\documentclass[journal,comsoc]{IEEEtran}
\usepackage[T1]{fontenc}

\usepackage{cite}

\usepackage{color}
\usepackage{graphicx} 
\usepackage{subcaption}
\usepackage{array}
\usepackage{verbatim}
\usepackage{makecell}
\usepackage{multirow}
\ifCLASSINFOpdf
\else
\fi
\usepackage{amsmath}
\usepackage[cmintegrals]{newtxmath}
\usepackage{url}

\begin{document}
%
\title{A Comprehensive Survey on Hybrid Communication for Internet of Nano-Things in Context of Body-Centric Communications}
%
%
%

\author{Ke~Yang, 
        Dadi Bi, 
        Yansha Deng, 
        Rui~Zhang, 
        M. Mahboob Ur Rahman, 
        Najah~Abu~Ali, 
        Muhammad Ali Imran, 
        Josep~M.~Jornet, 
        Qammer~H.~Abbasi, 
        and~Akram~Alomainy
        
   \thanks{     {\textit{(Ke Yang and Dadi Bi are both first authors.) (Corresponding author: Yansha Deng.)}}}
\thanks{K. Yang is with the School of Marine Science and Technology, Northwestern Polytechnical University, Xi'an, 710072, China (e-mail: k.yang@nwpu.edu.cn).}        
\thanks{D. Bi and Y. Deng are with the Department of Engineering, King's College London, London, WC2R 2LS, U.K. (e-mail:\{dadi.bi, yansha.deng\}@kcl.ac.uk).}
\thanks{R. Zhang is with the Department of Information and Electronic, Beijing Institute of Technology, Beijing, 100081, China (e-mail: rui.zhang@bit.edu.cn).}
\thanks{M. M. U. Rahman is with the Department of Electrical Engineering, Information Technology University, Lahore 54000, Pakistan  (e-mail: mahboob.rahman@itu.edu.pk).}
\thanks{N. A. Ali is with the College of Information Technology, United Arab
Emirates University, Al Ain, 15551, United Arab Emirates (e-mail: najah@uaeu.ac.ae).}
\thanks{M. A. Imran and Q. H. Abbasi are with the Department of Electronics
and Nanoscale Engineering, University of Glasgow, Glasgow, G12 8QQ, U.K. (e-mail: \{muhammad.Imran, qammer.abbasi\}@glasgow.ac.uk).}
\thanks{J. M. Jornet is with the Department of Electrical and Computer Engineering, Northeastern University, Boston, MA 02120, USA (e-mail: jmjornet@northeastern.edu).}
\thanks{A. Alomainy is with the School of Electronic Engineering and Computer Science, Queen Mary University of London, London, E1 4NS, U.K (e-mail: a.alomainy@qmul.ac.uk). }  }

\maketitle

\begin{abstract}
With the huge advancement of nanotechnology over the past years, the devices are shrinking into micro-scale, even nano-scale. Additionally, the Internet of nano-things (IoNTs) are generally regarded as the ultimate formation of the current sensor networks and the development of nanonetworks would be of great help to its fulfilment, which would be ubiquitous with numerous applications in all domains of life. However, the communication between the devices in such nanonetworks is still an open problem. Body-centric nanonetworks are believed to play an essential role in the practical application of IoNTs. BCNNs are also considered as domain specific like wireless sensor networks and always deployed on purpose to support a particular application. In these networks, electromagnetic and molecular communications are widely considered as two main promising paradigms and both follow their own development process. In this survey, the recent developments of these two paradigms are first illustrated in the aspects of applications, network structures, modulation techniques, coding techniques and security to then investigate the potential of hybrid communication paradigms. Meanwhile, the enabling technologies have been presented to apprehend the state-of-art with the discussion on the possibility of the hybrid technologies. Additionally, the inter-connectivity of electromagnetic and molecular body-centric nanonetworks is discussed. Afterwards, the related security issues of the proposed networks are discussed. Finally, the challenges and open research directions are presented.
\end{abstract}


%
\IEEEpeerreviewmaketitle

\section{Introduction to nanonetworks}

In upcoming years, the advancement in nanotechnology is expected to accelerate the development of integrated devices with the size ranging from one to a few hundred nano-meters \cite{Nano2008,EM2010}. With the aim of shrinking traditional machines and creating nano-devices with new functionality, nano-technologies have produced some novel nano-materials and nano-particles with novel behaviours and properties which are not observed at the microscopic level. The links and connectivity between nano-devices distributed through collaborative effort lead to the vision of nanonetworks, after the concept of nano-machine is proposed. The limited capabilities of nano-machines in terms of processing power, complexity and range of operations can be expanded by this collaborative communication. It is changing the paradigm from the Internet of Things (IoT) to Internet of Nano-Things (IoNTs) \cite{Nano2011} which shares the same development path as the nanonetworks. 

Communication between nano-machines in IoNTs can be set up through nano-mechanical, acoustic, chemical, electromagnetic (EM) and molecular communication approaches \cite{Nano2008}. Unfortunately, traditional communication technologies are not suitable mainly due to the limitations, such as size, complexity and energy consumption of transmitters, receivers and other components at nano-scale\cite{nishi2007scaling}; thus, novel and suitable communication techniques from physical layer to higher layers are required to develop for each paradigm.

The molecular and EM communication schemes are envisioned as two most promising paradigms and numerous researches have been done in these two paradigms. This review focuses on molecular and EM approaches and presents their backgrounds, applications, recent developments and challenges. We mainly present a comprehensive survey on the researches that have already been done to enable the communication in nanonetworks. Moreover, several aspects of the integration of nanonetworks have been identified. We propose to implement a hybrid communication taking advantage of both paradigms to enhance the communication performance and aim to broaden and realize more applications. The feasibility of the novel hybrid communication is discussed based on the requirements and enabling technologies from both micro and macro perspectives, and the open challenges are explored as a source if inspiration towards future developments of this inter-connectivity.

This paper provides a structured and comprehensive review on the recent literature on Body-Centric nanonetworks, an effectual foundation of IoNTs. The main contributions of this survey are summarized as follows.
\begin{itemize}
    \item The various applications are classified and summarized.
    \item The latest advancement in physical, link, MAC, network and application layers have been comprehensively reviewed in addition to security changes.
    \item The hybrid communication scheme by collaboratively employing EM-based nano-communication and molecular communication together is introduced.
    \item Open issues and challenges for such hybrid networks are introduced.
    \end{itemize}

The rest of the paper is organized as follows. Section \textrm{II} presents an overview of various communication paradigms, numerous applications and standardization. Section \textrm{III} discusses the general requirements and performance metrics of the envisioned body-centric nanonetworks, while  Section \textrm{IV} illustrates the enabling and concomitant technologies which would help the development of nanonetworks from EM and bio perspective, respectively. The architecture of the network and performance of EM and molecular communication is discussed in  Section \textrm{V} and  Section \textrm{VI}, respectively. The connectivity of both communication methods are discussed in  Section \textrm{VII}. In  Section \textrm{VIII}, the researches related to the security issues are discussed. In the end, the challenges and open problems are discussed with a brief conclusion.

\section{An Overview of nanonetworks}

According to Feynman, there is plenty of room at the bottom \cite{feynman1960there}.
Based on such statement and the considerable development of nano-technology, Prof. Metin Sitti has proposed that in the near future the network would go down to the nano-scale if the nano robots and molecular machine are adopted as its elements \cite{sitti2015biomedical}. Thus, the concept of nano-networks was proposed. However, the connection between nano-devices in such networks would be a challenge, leading to the study on nano-communication \cite{bush2010nanoscale,akyildiz2008nanonetworks}. Therefore, nano-communication can be defined as the communication between nano-devices where the communication principles should be novel and modified to meet the demands in the nano-world. To make it clearer, four requirements are summarized in IEEE P1906.1 \cite{ieeedraft} in the aspects of components, system structure, communication principles and \textit{etc.},

\subsection{Nano-communication paradigms}
To make the network work well, the communication between the nano-devices needs to be linked. In \cite{akyildiz2008nanonetworks}, nano-communication is studied in two scenarios: (1) Communication between the nano-devices to the micro/macro-system, and (2) Communication between nano-devices. Furthermore, molecular, electromagnetic, acoustic, nano-mechanical communication can be modified to nano-networks \cite{andrew2000nanomedicine}, summarized in our previous work in \cite{nano2016}.  Based on the burgeoning of the nanotechnology, a fresh model of mechanical communication, \textit{i.e.} touch communication (TouchCom), was also proposed in \cite{chen2015touch}, where bunches of nano-robots were adopted to play as the message carriers. In TouchCom, transient microbots (TMs) \cite{hwang2012physically, martel2009flagellated, martel2009mri} were used to carry the drug particles and they are controlled and guided by the external macro-unit (MAU) \cite{khalil2013magnetic, chen2014conceptual}. These TMs would stay in the body for some time whose pathway is the channel and the operations of loading and unloading of drugs can be treated as the transmitting and receiving process. The channel model of TouchCom could be described by the propagation delay, loss of the signal strength in the aspects of the angular/delay spectra\cite{chen2015touch}. A simulation tool was also introduced to characterize the action of the nano-robots in the blood vessel \cite{chen2014conceptual}.

\subsection{Applications of nanonetworks}

Nano-communication spans a wide area such as military, ubiquitous health care, sport, entertainment and many other areas, detailed description of which has been summarized and classified in \cite{akyildiz2008nanonetworks}, shown in Table \ref{tab:overviewofapplications}. The main characteristic in all applications is to improve people's life quality and nanonetworks are generally believed as the perfect candidate for bio-medical fields due to bio-compatibility, bio-stability and its dimension. Generally, the applications are classified into two general categories, medical and non-medical as below.

\begin{table*}[tb]
    \centering
    \caption{Summaries of the pictured applications \cite{sitti2015biomedical,akyildiz2008nanonetworks}}
    \label{tab:overviewofapplications}
    \resizebox{\textwidth}{!}{
        \begin{tabular}{clccc}
        \hline
        \multicolumn{2}{c}{Biomedical \cite{freitas2005nanotechnology}}                         & Environmental                     & Industrial                                & Military                                                    \\ \hline
                                         & $\bullet$ Active Visual Imaging for Disease Diagnosis \\ &\cite{liao2010indications} \cite{nakamura2008capsule} \cite{pan2011swallowable} \cite{fluckiger2007ultrasound} \cite{kim2008noninvasive} &                                   &                                           &                                                             \\
        \multirow{-2}{*}{Health Monitor} & $\bullet$ Mobile Sensing for Disease Diagnosis \\ & \cite{ergeneman2008magnetically} \cite{dubach2007fluorescent} \cite{li2003cholesterol} \cite{tallury2010nanobioimaging}       & \multirow{-2}{*}{Bio-Degradation \cite{akyildiz2010electromagnetic}} & \multirow{-2}{*}{Product Quality Control \cite{aylott2003optical}} & \multirow{-2}{*}{Nuclear, Biological and Chemical Defences \cite{avouris2000carbon}} \\ \hline \hline
                                         & $\bullet$ Tissue Engineering \\ & \cite{tasoglu2014untethered} \cite{fox2014use} \cite{kim2013fabrication}                         & \multicolumn{1}{l}{}              & \multicolumn{1}{l}{}                      & \multicolumn{1}{l}{}                                        \\
                                         & $\bullet$ Bio-Hybrid Implant\\ & \cite{drexler1992nanosystems} \cite{freitas2005nanomedicine}                         &                                   &                                           &                                                             \\
                                         & $\bullet$ Targeted Therapy$/$Drug Delivery \\ & \cite{fernandez2007magnetic} \cite{freitas2006pharmacytes} \cite{timko2010remotely} \cite{yim2012shape} \cite{carlsen2014bio}           &                                   &                                           &                                                             \\
                                         & $\bullet$ Cell Manipulation \\ & \cite{chen2005development} \cite{liao2010indications} \cite{steager2013automated} \cite{kawahara2013chip} \cite{kim2009microengineered}                          &                                   &                                           &                                                             \\
        \multirow{-10}{*}{Therapy}        & $\bullet$ Minimally Invasive Surgery \\ & \cite{kong2005rotational} \cite{miloro2012removing} \cite{yim2014biopsy}                 & \multirow{-10}{*}{Bio-Control \cite{heil2008long} \cite{pieterse2007plant} \cite{han2008molecular} }     & \multirow{-10}{*}{Intelligent Office \cite{akyildiz2010internet}}      & \multirow{-10}{*}{Nano-Fictionalized Equipment \cite{andrew2000nanomedicine} } \\
        \hline
        \end{tabular}
    }
    \end{table*}

	\begin{figure}[!t]
	\centering
	
	\begin{subfigure}{0.5\textwidth}
	\centering
	\includegraphics[width=0.6\textwidth]{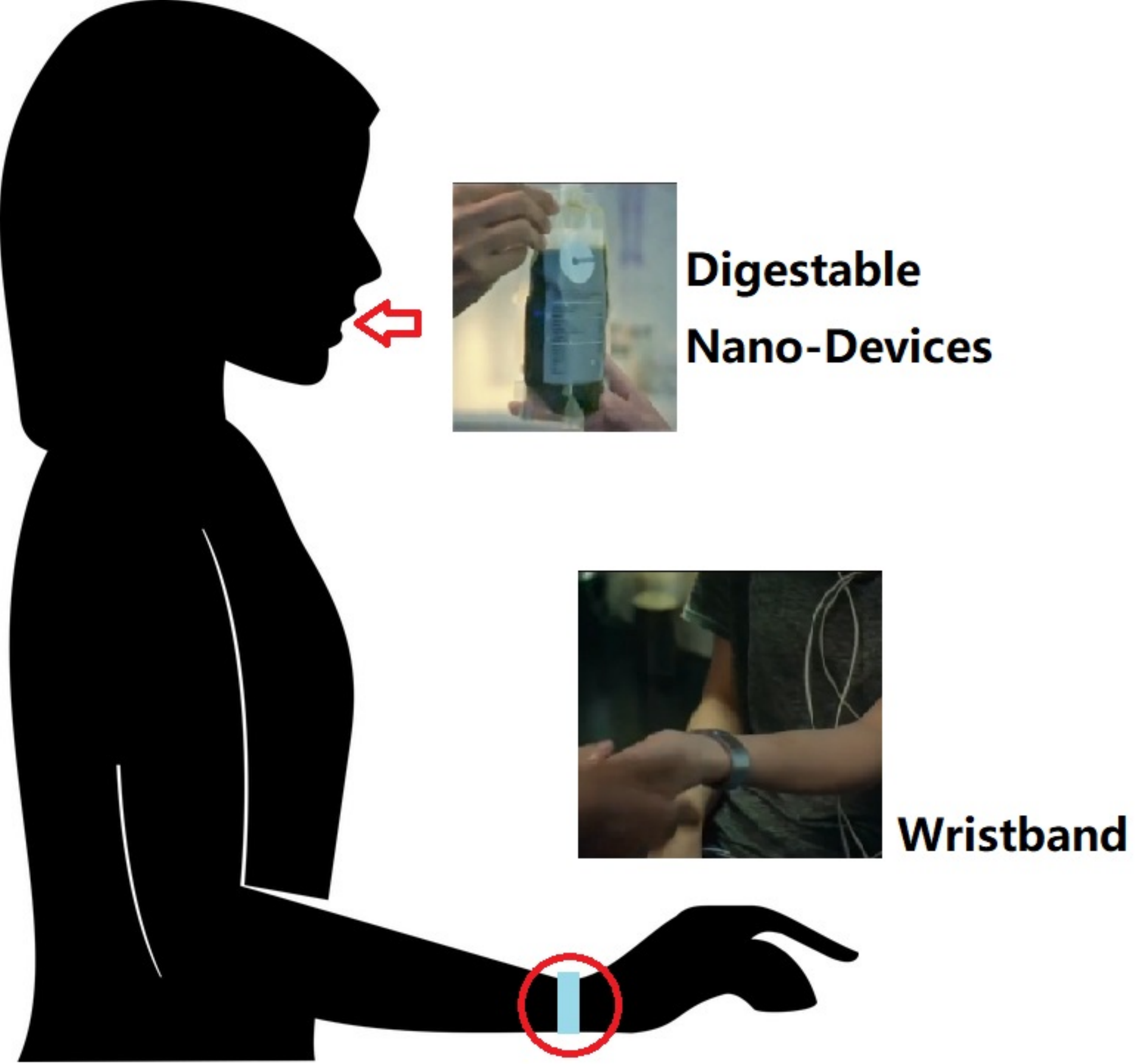}
	\caption{Scheme of the procedure}
	\label{ky_fig:circledisplay}
	\end{subfigure}
	
	\begin{subfigure}{0.5\textwidth}
	\centering
	\includegraphics[width=0.8\textwidth]{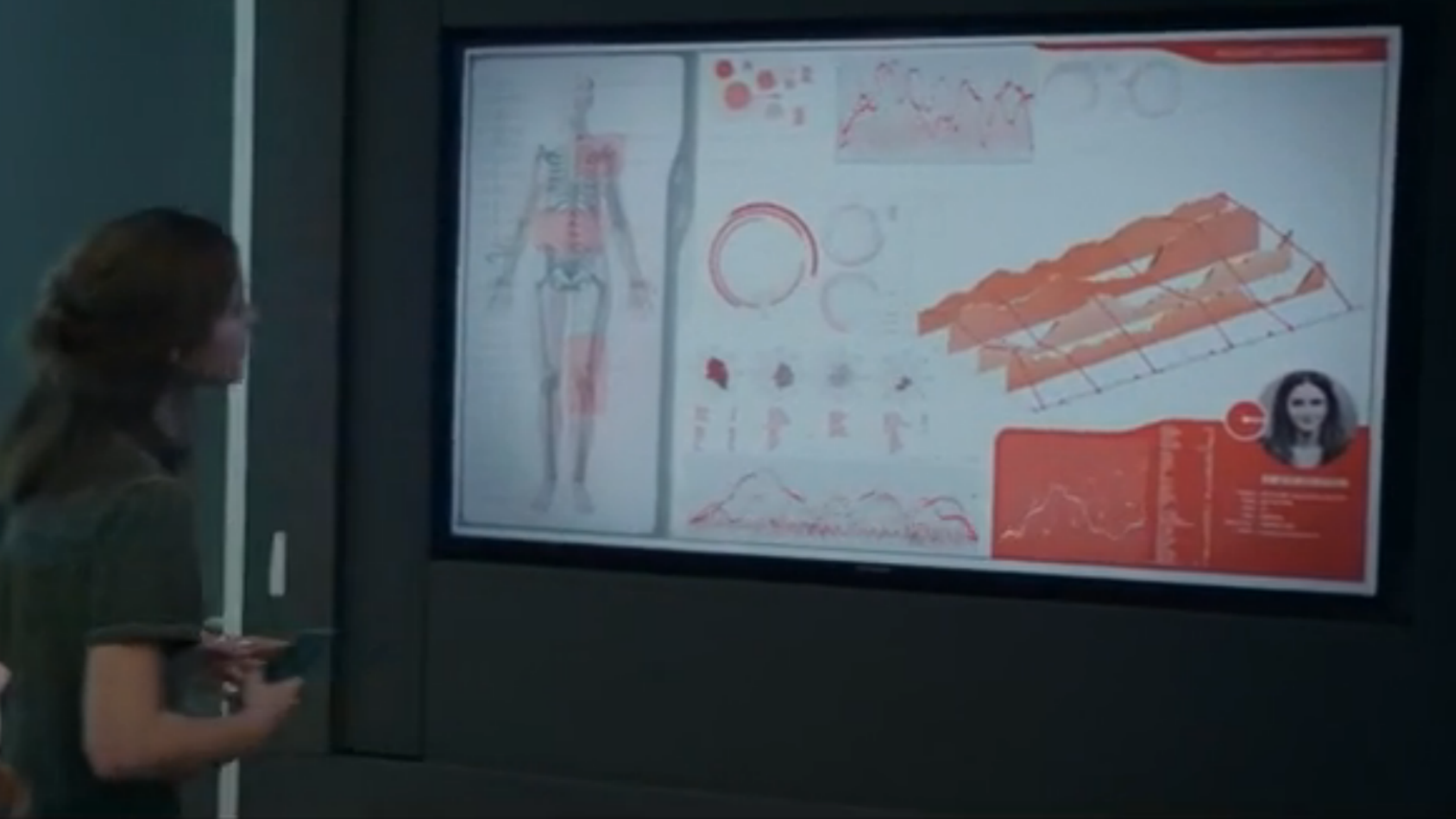}
	\caption{Capture of the movie showing how the vital health information was shown}
	\label{ky_fig:vitalsigns}
	\end{subfigure}
	
	\caption{Health monitoring system described at the film of \textit{The Circle} (from 34:35 to 35:55)}
	\label{ky_fig:healthmonitor}
	\end{figure}

\subsubsection{Medical Applications} \label{medicap_app}  
There are many biomedical application in literature, e.g, intraocular pressure (IOP) for vision  \cite{sitti2015biomedical} and nano robotos for cancer cells in  \cite{santagati2014opto}. Moreover, nanonetworks will monitor the body status in real time and some nano-devices can be used as tissue substitutes, i.e., bio-hybrid implants. In the following, we present two interesting examples that come from movies, which shows the limitless possibilities of nano network medical applications.

\paragraph{Health-Monitoring}

In the movie \textit{The Circle}, an example of the health-monitor system which is installed in the body of the lead actress May has been displayed. The whole system consists of two parts: digestible nano-sensors and a wristband. At first, the doctor asked May to drink a bag of green solution with the nano-sensors in and then gave her a wristband which should be worn all the time, shown in Fig. \ref{ky_fig:circledisplay}. The medium band would sync up with the sensors May has swallowed while both equipments would collect data of the heart rate, blood pressure, cholesterol, sleep duration, sleep quality, digestive efficiency, and so on. The capture of the movie in Fig. \ref{ky_fig:vitalsigns} shows the related information on the wall. Through the wristband, all the data can be stored anywhere May wants. Also, all the data would be shared with the related people, like the doctor or the nutritionist.

\paragraph{Real-Time Detection}

\begin{figure}[!t]
\centering
\includegraphics[width=0.5\textwidth]{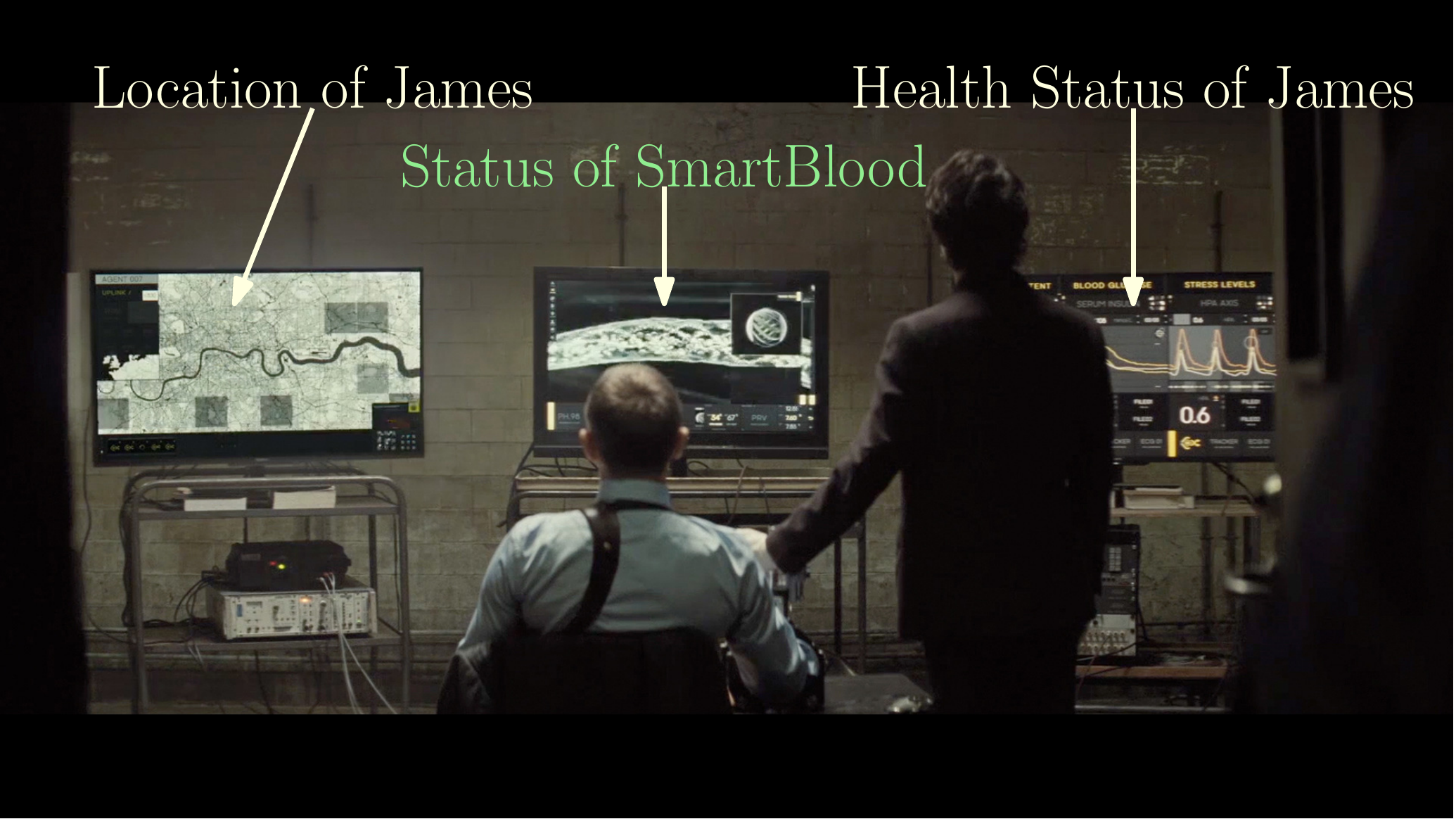}
\caption{Injection of \textit{SmartBlood} James Bond at the file of \textit{007: Spectre} (from 25:12 to 25:37)}
\label{ky_fig:smartblood}
\end{figure}

In the movie of \textit{007:~Spectre}, a technology called Smart Blood was illustrated, which is a bunch of nano-machines/micro-chips capable of tracking Mr. Bond's movement in the field. They were injected into Bond's blood system, and the institute would monitor Bond's vital signs from anywhere on the planet, shown in Fig. \ref{ky_fig:smartblood}. It is not just a scientific idea in the movie because several researchers are working on various kinds of injectable substances that can identify cancer cells. The folks at Seattle-based Blaze Bio-science are among the pioneers.

\begin{figure}[!htpb]
\centering
\includegraphics[width=0.45\textwidth]{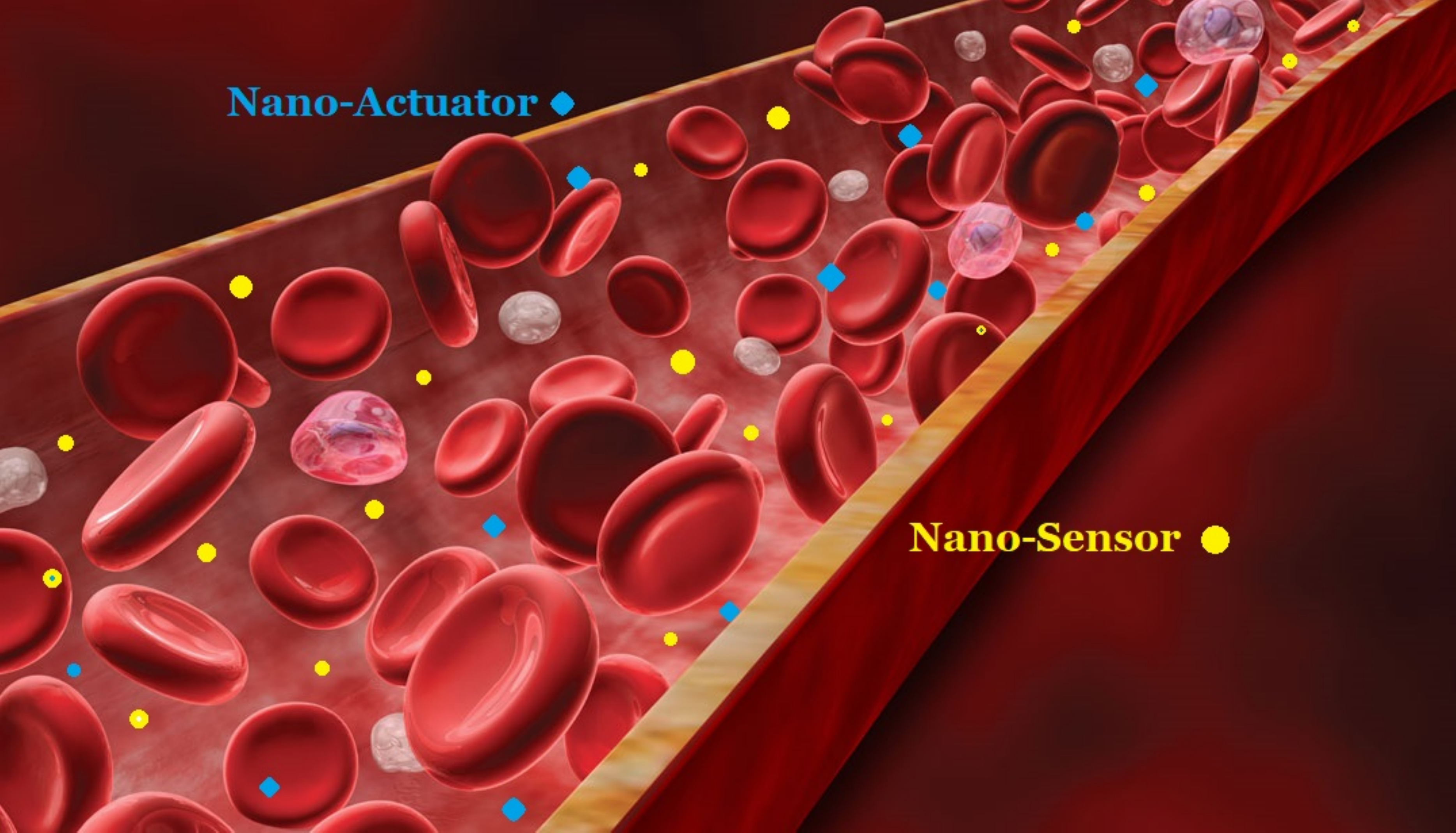}
\caption{Nanonetwork for drug delivery}
\label{ky_fig:blood gluecose detection}
\end{figure}

\paragraph{Drug Delivery} 
It is believed that the nanonetworks can not only sense the information but also make some actions when needed. The most trustworthy application would go for real-time glucose control. The nano-sensors spreading in blood vessels can monitor the glucose level; at the same time, the nano-machines could release the insulin to regulate the glucose level (shown in Fig. \ref{ky_fig:blood gluecose detection}). With such technologies, people with diabetes would not need to needle themselves and inject the medicine in public which would cause the embarrassment and infection if the operation is not correct. Also, the signal can also be sent to related people through wearable devices or smart phones to let them help the patients build a healthy habit.

\subsubsection{Non-Medical Applications}

\begin{figure}[!t]
	\centering
	\begin{subfigure}{0.49\textwidth}
	\centering
	\includegraphics[width=0.5\textwidth]{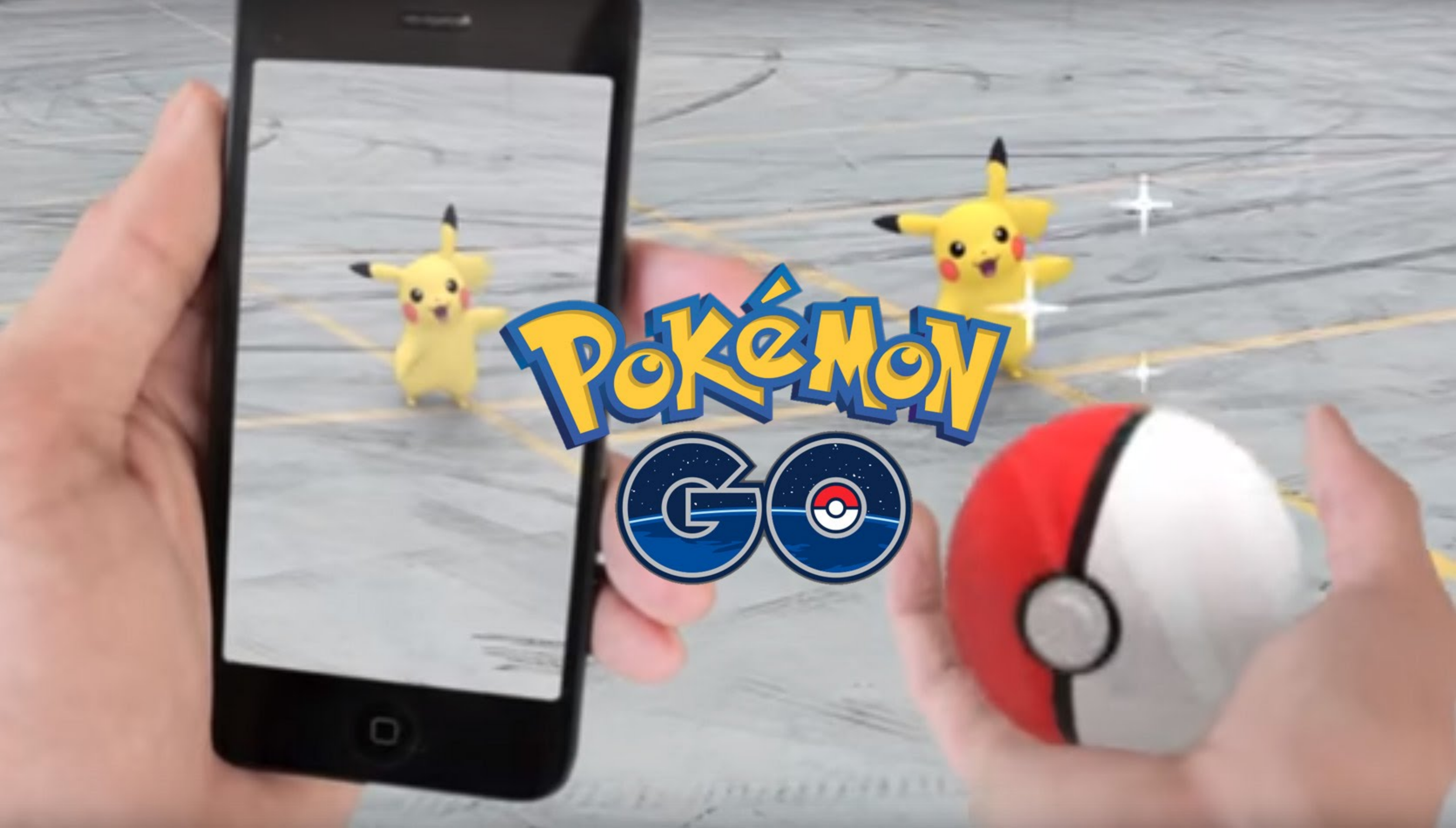}
	\caption{current technology for visual/augmented reality}
	\label{ky_fig:realmon}
	\end{subfigure}
	
	\begin{subfigure}{0.49\textwidth}
	\centering
	\includegraphics[width=0.5\textwidth]{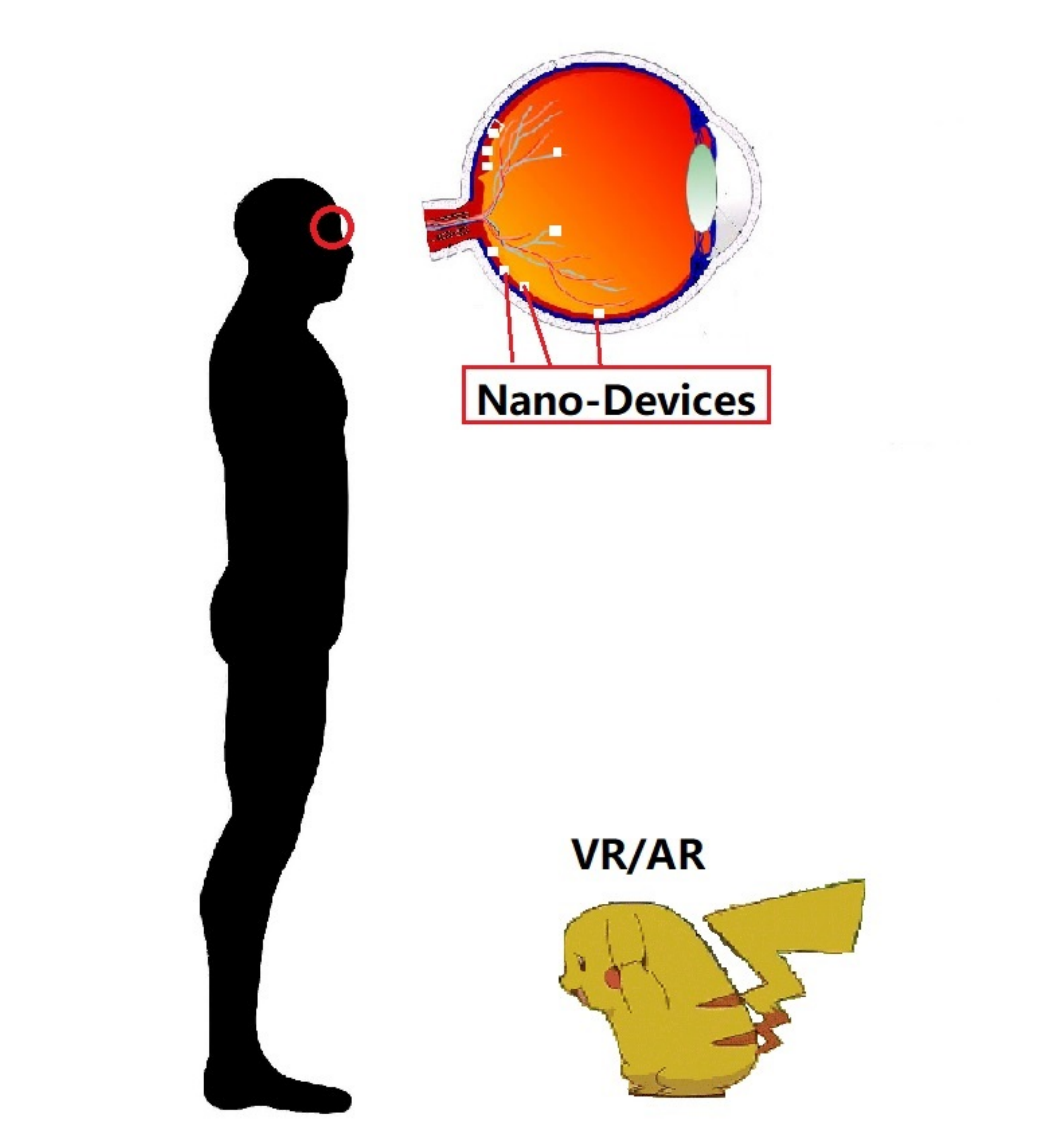}
	\caption{Envisioned scenario for future visual/augmented reality}
	\label{ky_fig:envisionmon}
	\end{subfigure}
	
	\caption{Nanonetwork for entertainment}
	\label{fig:vr}
	\end{figure}

\paragraph{Entertainment (VR/AR)} 
Currently, the realization of the visual/augmented reality requires the help of external devices like smartphones, shown in Fig. \ref{ky_fig:realmon} which is bulky and not convenient. If the nano-devices are spreading over the eyes, near the retina, they would help people see things as required. At the same time, the nano-machines spreading over the body would excite different parts of human to make the experience real. Take the Pokemon shown in Fig. \ref{ky_fig:envisionmon} as an example, the nano-machines in the eyes would help people see the monster in real world. If you want to catch the monster, all you need to do is just throwing you arm and the sensors in the arm would capture this action and judge if it is the right path and strength to capture the monster. If the monster fight, such as the shock generated by Pikachu, the nano-machines on the skin would cause some itches or aches to you. It is widely thought that such new technologies would cause radical changes to the current game experiences and would also help people gain the experiences they never have.

\begin{figure}[!t]
\centering
\includegraphics[width=0.3\textwidth]{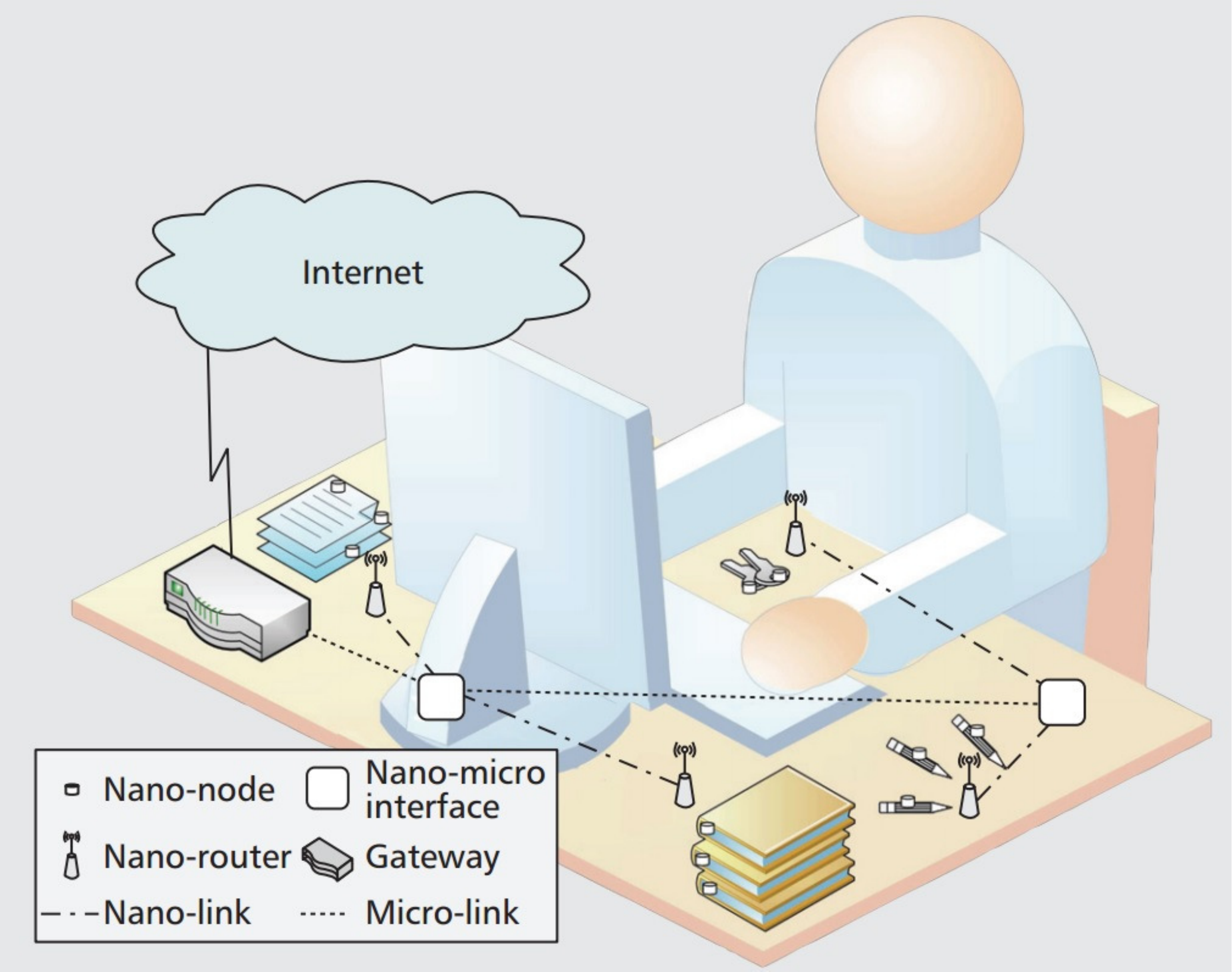}
\caption{Network architecture of the e-office \cite{akyildiz2010internet}}
\label{fig:eoffice}
\end{figure}

\paragraph{E-Environment} 
The illustration of the e-office is shown in Fig. \ref{fig:eoffice}. Every elements spreading over the office or the internal components are nano devices permanently connecting to the Internet. Thus, the locations and statuses of all belongings can be tracked effortlessly. Furthermore, by analysing all the information collected by nanonetworks of the office, actuators can make the working environment pleasant and intelligent.

\paragraph{Agriculture/Industry Monitoring} 

\begin{figure} [!t]
\centering

\begin{subfigure}{0.5\textwidth}
\centering
\includegraphics[width=0.55\textwidth]{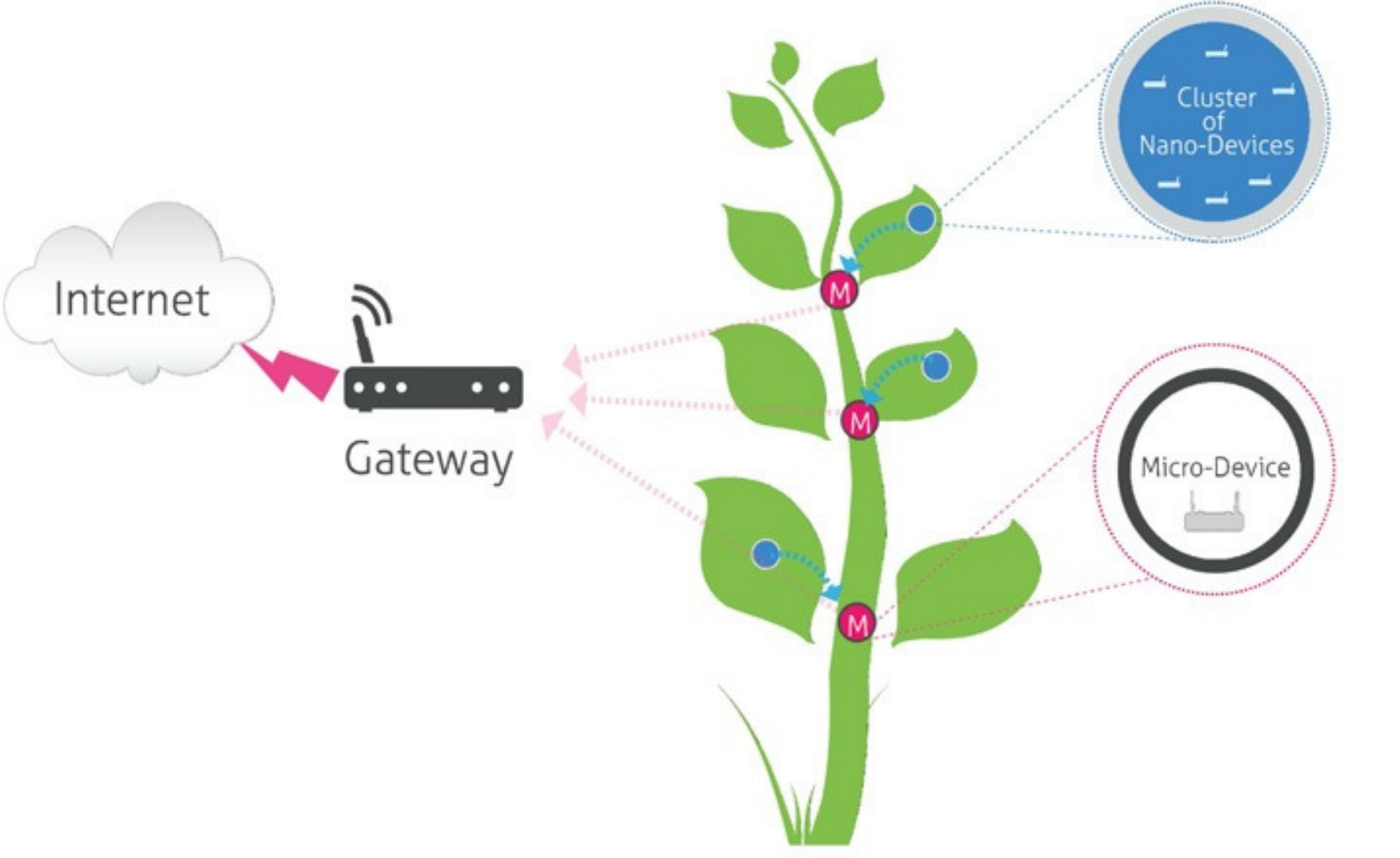}
\caption{Network architecture}
\label{fig:strc_PlntMnt}
\end{subfigure}

\begin{subfigure}{0.5\textwidth}
\centering
\includegraphics[width=0.5\textwidth]{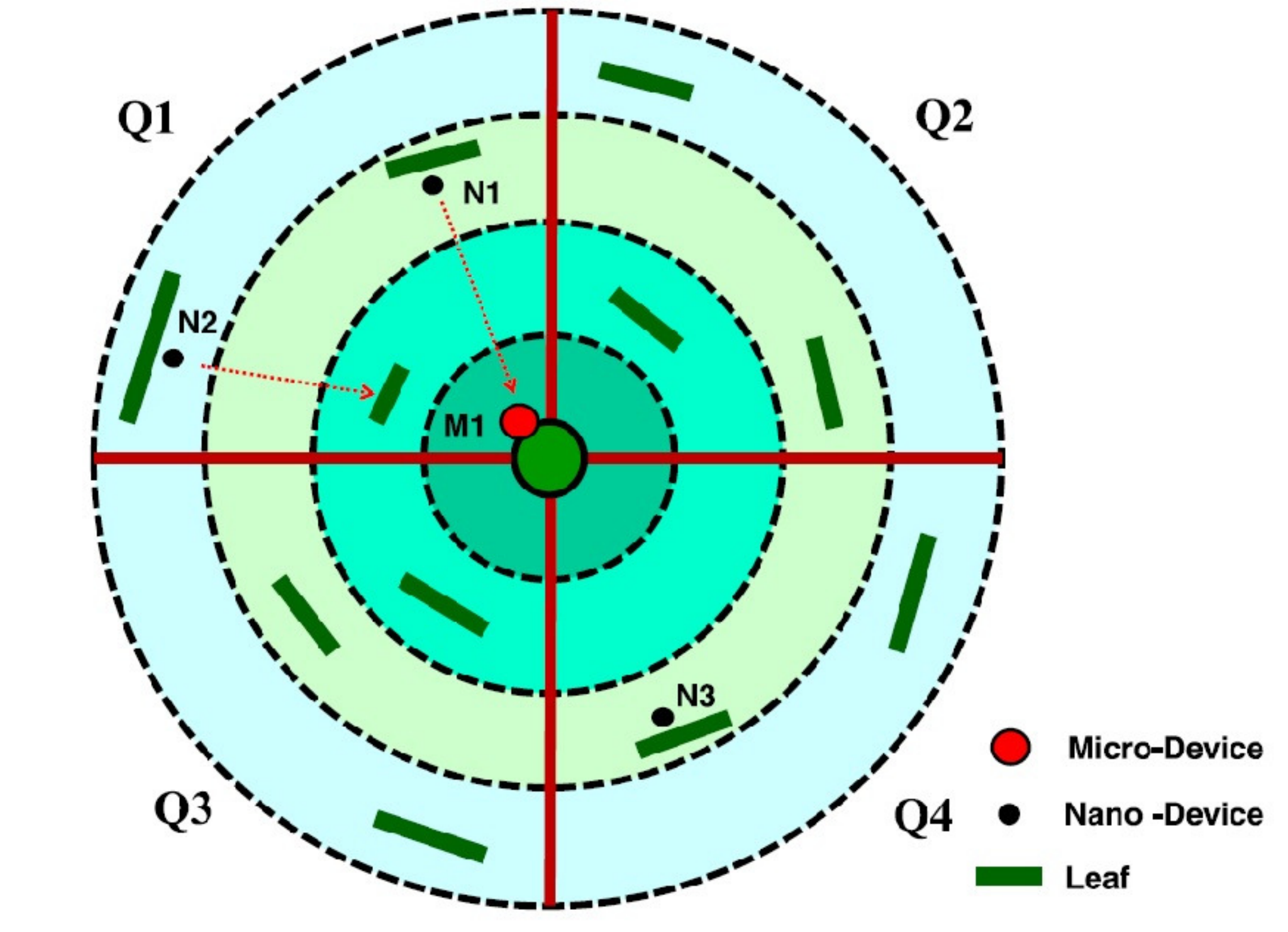}
\caption{Details of the nodes distribution}
\label{fig:strc_PlntMnt_details}
\end{subfigure}

\caption{nanonetworks for plant monitoring \cite{afsharinejad2015dynamic}}
\label{fig:strc_PlntMnt1}
\end{figure}

Fig. \ref{fig:strc_PlntMnt1} shows an example of using nanonetworks for crop-monitor \cite{afsharinejad2015dynamic}. Since the plants would release typical chemical compound which would be used to analyse the environment conditions and plant growth condition. The structure of such monitor network is described in \cite{afsharinejad2015dynamic}, shown in Fig. \ref{fig:strc_PlntMnt}. It is said that such systems can not only monitor growth status of the plants but also analyse the underground soil and air conditions which can be used as a chemical defence system.

\subsection{Standardization of nanonetworks}
\label{standard}
To make nanonetworks function well, the IEEE has built a
standard development project: IEEE P1906.1/Draft 1.0 Recommended Practice for Nano-scale and Molecular Communication Framework \cite{ieeedraft}, leading by S.F. Bush. The initial ballot was done on Jan. 2 2015 and then materials related to SBML (Systems Biology Markup Language) was added to make the draft complete in August, which was approved by IEEE RevCom later in 2015. Currently, \textit{IEEE 1906.1.1 - Standard Data Model for Nano-scale Communication Systems} is proposed to define a network management and configuration data model for nanonetworks, written in YANG Model (Yet Another Next Generation).

The IEEE P1906.1/Draft 1.0 specifies a framework for networks rather than protocols or layering. The standard defines the framework through the use of the term, ``component'', to limit any alluding to layering or a protocol stack. The provided services by the components remain an integral part of the process for the framework. As such, protocols are necessary to manage this process. The nano-scale communication network is comprised of five fundamental components necessary for the deployment of the network: the message carrier, motion, field, perturbation, and specificity. 

\begin{enumerate}
	\item The \textbf{Message Carrier} may take the form of a particle or a wave and is defined as the physical entity that carries a message.  
	\item The \textbf{Motion Component} provides force that enables the message carrier to move. Motion provides the necessary potential to transport information through a communication channel.
	\item The \textbf{Field Component} guides the message carrier. For example, an internal implementation includes swarm motion or flocking behavior while external implemen-tations may include non-turbulent fluid flow, EM field, a chemical gradient released to guide the movement of bacteria, or molecular motors guided by micro-tubules.
	\item \textbf{Perturbation} provides the service of varying Message Carriers as needed to represent a signal. It functions like modulation in telecommunications. Perturbation can be achieved via varying signals based on the number of received message carriers, controlled dense vs. sparse concentrations of molecules, simple on vs. sparse concentrations of molecules, simple on vs. off flow of signal molecules. It uses different types of message carriers, modifying the conformation of molecules to represent multiple states for the component that provides controlled change to create a signal.
	\item \textbf{Specificity} provides the function of reception of a message carrier by a target. It is analogous to addressing in classical communication systems. Specificity can be seen in the shape of a molecule or its affinity to a target, such as complementary DNA for hybridization.
\end{enumerate}

The framework also defines the relationships and how each component is interfaced in relation to the other components. This will allow for a broader and more encompassing definition for networking when compared to the classical networking protocol stack and OSI layering system. Table \ref{tab:table1} and \ref{tab:table2} show the function of nanonetworks components and map the relationship between the IEEE P1906.1 framework and the classical networking protocol stack, respectively. 
\begin{table*}[!t]
    \centering
    \caption{An example of nanoscale communication network components [IEEE P1906.1].}
    \begin{tabular}{p{55pt}|p{80pt}|p{170pt}|p{110pt}}
         \hline
         Layer name & Explanation & Example (molecular) & Example (nanotube/terahertz) \\ \hline
         Specificity & Correctly detect true versus false messages & Shape or affinity of molecule to a particular target complementary DNA for hybridization, \textit{etc}. & Antenna aperture, resonant frequency, impedance match\\
         \hline
         Perturbation & Vary concentration or motion as needed for signal (shockwave) & Dense versus sparse concentrations of molecules, on versus off flow of signal molecules or motors, conformational changes in molecules, \textit{etc}. & Amplitude, frequency, or phase modulation \\ \hline
         Field & Organized flow direction & Flowing liquid applied EM field, motors attached to microtubules, concentration gradient of chemical molecules, swarm motion, \textit{etc}. & Omni or directed with multiple CNTs \\ \hline
         Motion & Potential communication channel in the wild (semi-random) & Molecules diffusing through liquid, unattached molecular motors, Brownian motion, self-propelled motion, \textit{etc}. & Wave propagation and phase velocity \\ \hline
         Message Carrier & Mass and energy & Molecule chain, \textit{etc}. & EM wave \\
        \hline
    \end{tabular}
    \label{tab:table1}
\end{table*}
\begin{table*}
\caption{An example of OSI to nanoscale communication network mapping [IEEE P1906.1].}
\begin{tabular}{l|l|l|l|l}
\hline
OSI model                                               & \multicolumn{3}{l}{\begin{tabular}[c]{@{}l@{}}IEEE 1906 component mapping\end{tabular}}    & Explanation                                                                                                                                                                      \\ \hline
Application                                             &                                                          &                        &              & \begin{tabular}[c]{@{}l@{}}No 1906 component\end{tabular}                                                                                                                   \\ \hline
Presentation                                            &                                                               &                        &              & \begin{tabular}[c]{@{}l@{}}No 1906 component\end{tabular}                                                                                                                   \\ \hline
Session                                                 &                                                               &                        &              & \begin{tabular}[c]{@{}l@{}}No 1906 component\end{tabular}                                                                                                                   \\ \hline
Transport                                               &                                                               &                        &              & \begin{tabular}[c]{@{}l@{}}No 1906 component\end{tabular}                                                                                                       \\ \hline
Network                                                 &                                                               & \multirow{1.5}{*}{Field} &              & \begin{tabular}[c]{@{}l@{}}Field may enable Message Carrier transport across multiple nodes\end{tabular}                                      \\ 
\begin{tabular}[c]{@{}l@{}}Data Link\end{tabular} & Specificity                                                   &                        &              & \begin{tabular}[c]{@{}l@{}}Motion, enhanced by Field and Specificity, enable Message Carrier to reach next-hop node\end{tabular} \\ 
Physical                                                & \begin{tabular}[c]{@{}l@{}}Message Carrier\end{tabular} & \multirow{1.5}{*}{Motion}                  & Perturbation & \begin{tabular}[c]{@{}l@{}}Perturbation creates the signal transported by the Message Carrier using Motion\end{tabular} \\ \hline                  
\end{tabular}
\label{tab:table2}
\end{table*}

\begin{figure}[t]
	\centering
	\includegraphics[width=0.4\textwidth]{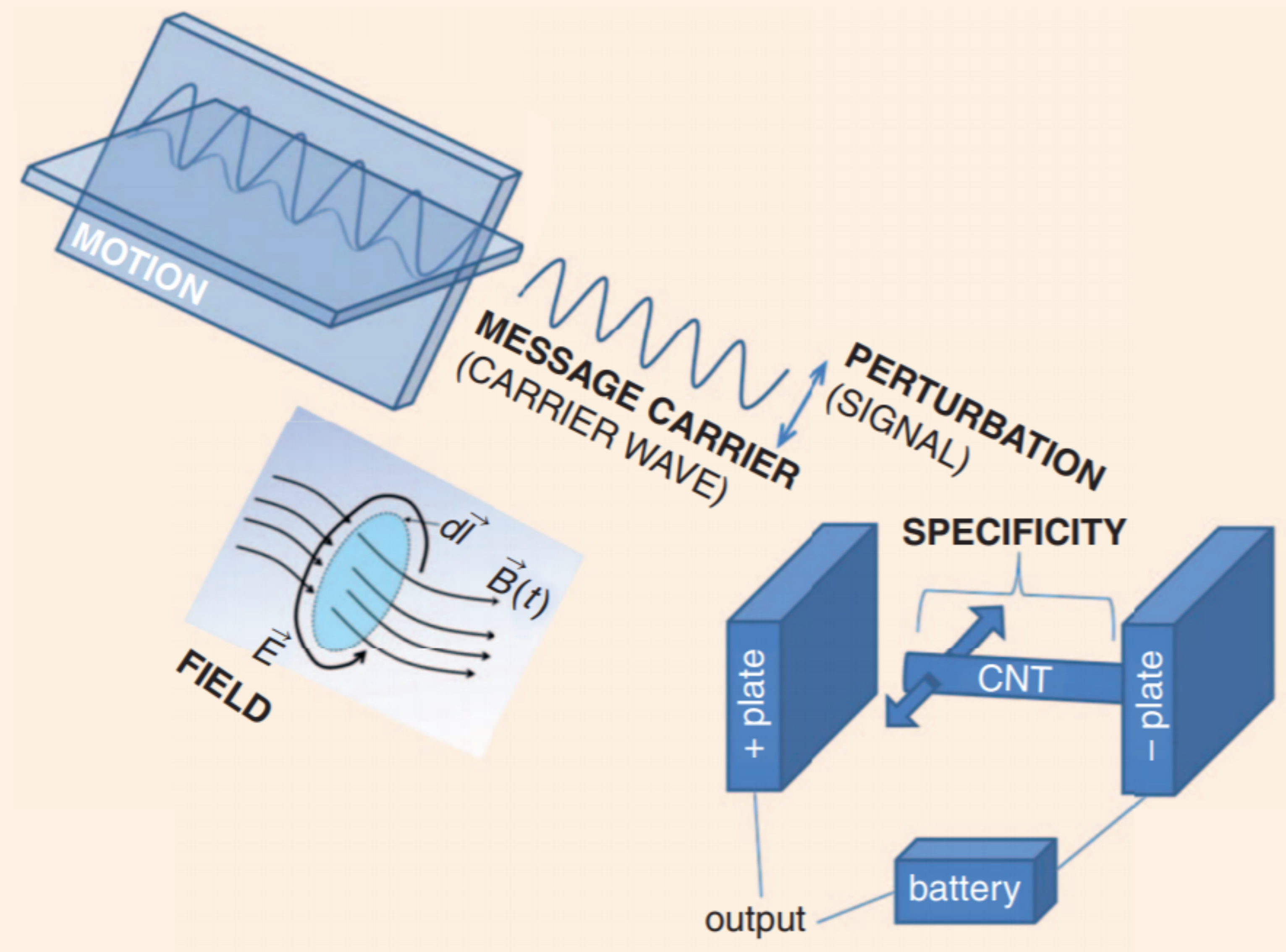}
	\caption{Message Carrier: CNT radio \cite{Bush}}
	\label{figure:najah_figure1}
\end{figure}
\begin{figure}[t]
	\centering
	\includegraphics[width=0.4\textwidth]{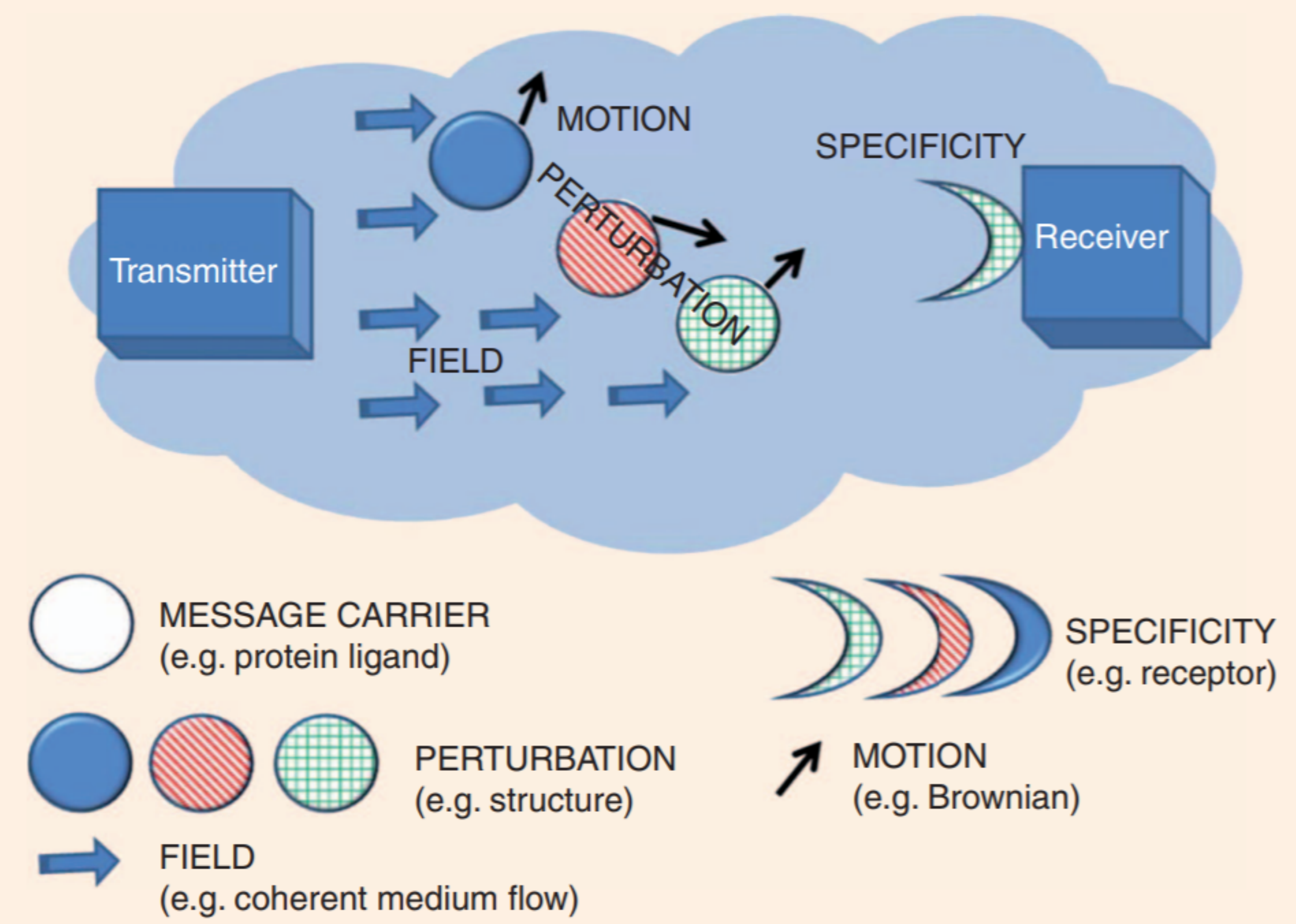}
	\caption{Message carrier in molecular communication \cite{Bush}}
	\label{figure:najah_figure2}
\end{figure}

An instant of the framework in an active network may include the message carrier component that transports a message, as shown in Figs. \ref{figure:najah_figure1} and \ref{figure:najah_figure2} \cite{Bush}. The specificity component provides message addressing, which aids in message delivery to the correct receiver. The perturbation component aids in signal formation by applying necessary variations to motion or concentration that lead to the recognition by the required target or receiver. Finally, the message is moved across the network physically through motion, and the field acts as a directional vector of motion that guides the message towards the target or receiver.. 

Other elements of a nanonetwork are also defined by the framework, such as the nanonetworks interface to micro/macro classical networks and the relay; details are given in Sec. \ref{hy_com}) to micro/macro classical networks and the relay, but did not discuss the required number of interfaces to the classical networks or the number of components required for accurate detection of an event or performing a specific function such as drug delivery. These issues are left for academia and industry to provide innovative solutions.

\section{Requirements and Performance metrics of Body-Centric Nanonetworks}

\subsection{EM-Based Body-Centric Nanonetworks}
\subsubsection{Achievable Information Rates}

\begin{figure}[!t]
	\centering
	\includegraphics[width=0.5\textwidth]{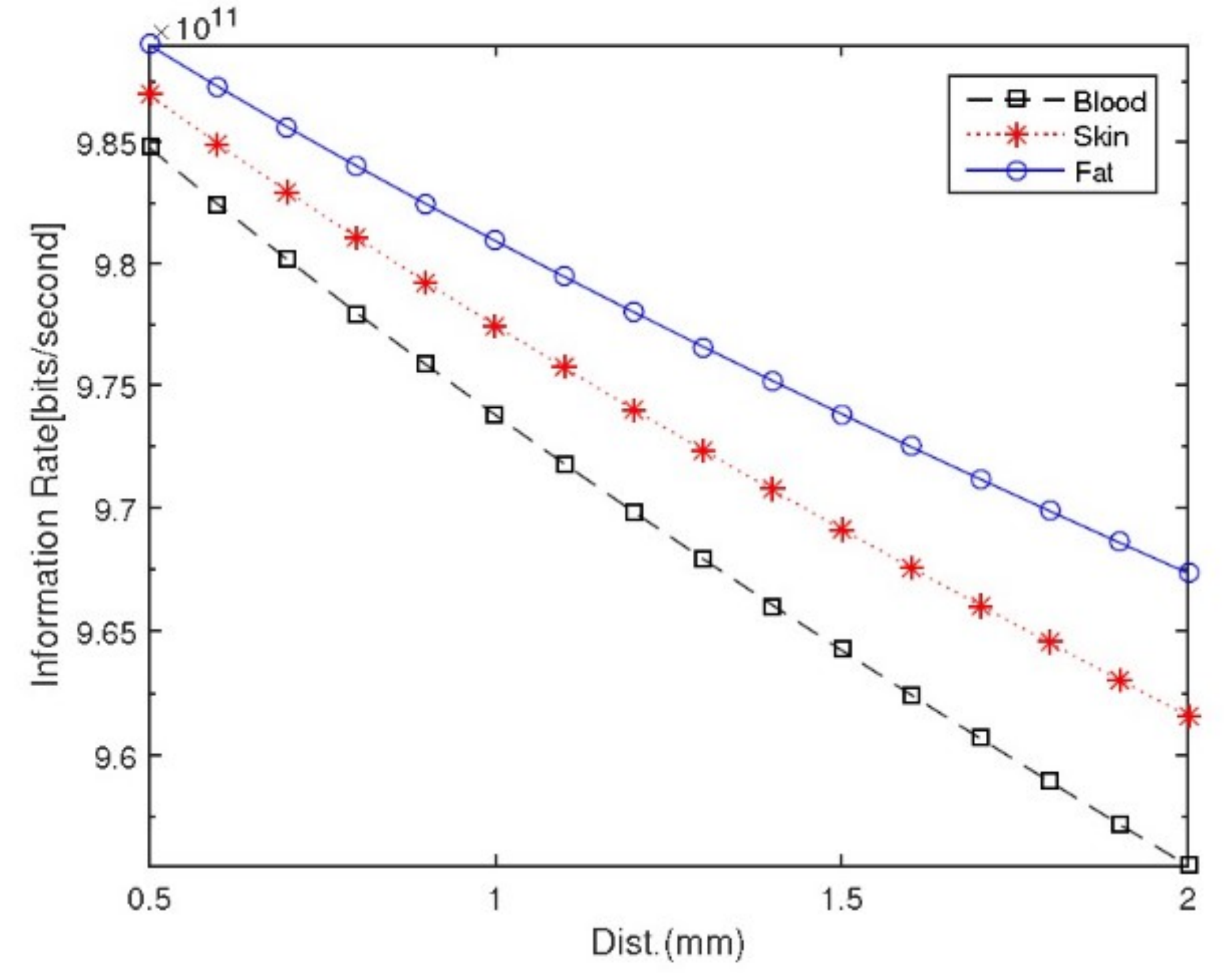}
	\caption{The trade-off between Information rate and transmission distance for three different human tissues {{\cite{qammer4}}}.}
	\label{fig:qm_eoffice1}
\end{figure}

The maximum achievable information rate, $I{R_{{\rm{max}}(sym)}}$, with the unit of bit/symbol based on a specific modulation scheme in a communication system has been defined as \cite{shannon1948a}:

\begin{equation}
\label{eq:qammer1}
I{R_{{\rm{max}}(sym)}} = \mathop {\max }\limits_x \left\{ {\rm{H}\left( X \right) - H(X|Y)} \right\},
\end{equation}
where X and Y denote the message sent by the transmitter and its noisy version at the receiver, respectively. Here, H(X) represents the entropy of message X, while H(X|Y) is the conditional entropy of X given Y. Represent the transmitted information over the asymmetric THz band channel without coding as a discrete binary random variable, ${x_0}$ and ${x_1}$; then, H(X) is given as \cite{qammer1}:
\begin{equation}
\label{eq:qammer2}
{\rm{H}}\left( {\rm{X}} \right) =  - \mathop \sum \limits_{m = 0}^1 \left\{ {{p_X}({x_m}) \cdot lo{g_2}{P_X}({x_m})} \right\},
\end{equation}
where ${p_X}({x_m})$ indicates the probability of transmitted symbol ${x_0}$ named as silence and ${x_1}$ named as pulse.
Assuming Additive Coloured Gaussian Noise (ACGN) \cite{qammer2} at the receiver, and a Binary Asymmetric Channel (BAC) with Y being a discrete random variable, the information rate (in bits/second) is given as 
\cite{low2014}:
\begin{equation}
\label{eq:qammer3}
I{R_{{\rm{max}}(sec)}} = \frac{B}{\beta }I{R_{{\rm{max}}(sym)}},
\end{equation}
where B represents the bandwidth of channel. $\beta$ is the ratio of the symbol interval ${T_s}$ and the pulse length ${T_p}$. And the rate of the symbols transmitted is defined as ${\rm{R}} = {\raise0.7ex\hbox{$1$} \!\mathord{\left/{\vphantom{1{{T_s}}}}\right.\kern-\nulldelimiterspace}\!\lower0.7ex\hbox{${{T_s}}$}} ={\raise0.7ex\hbox{$1$} \!\mathord{\left/{\vphantom {1{(\beta{T_p})}}}\right.\kern-\nulldelimiterspace}\!\lower0.7ex\hbox{${(\beta {T_p})}$}}$.
Note that the requirements on the transceiver can be greatly relaxed by reducing the single-user rate to increase $\beta$. Fig. \ref{fig:qm_eoffice1} studies the trade-off between the information rate and the transmission distance, for three different human body tissues (with an EM channel of bandwidth 1 THz {\cite{qammer4}}. 

\subsubsection{Bit  Error Rate}
Since EM waves propagate through the frequency-dependent materials inside the human body, the operating frequency has an important effect on the communication channel. \cite{guo2016intra} shows that the scattering from cells is the major phenomenon affecting the propagation of EM waves at optical frequencies inside the human body.\cite{joint2016} does the error analysis (at the physical layer and link layer) of an EM system operating in THz band.

\subsubsection{Symbol Error Rate}
\cite{rahm2013thz} studies different types of modulators capable of setting the amplitude or phase of the THz wave. A meta-material-based modulator was employed to control the phase of THz wave in \cite{chen2009a}. \cite{singh2016graphene-based} proposes and validates an analytic model for the plasmonic phase modulator that starts from the dynamic complex conductivity of graphene. By applying the model, the symbol error rate performance of the plasmonic modulator is studied when it is utilized to implement an M-array phase shift keying modulation. 

\subsection{MC-Based Body-Centric Nanonetworks}  
\subsubsection{Achievable Information Rates}

The discussion of the performance limits of the MC-based nanonetworks in terms of achievable information rates was first initiated by \cite{alfano2006on}. 
Later, Eckford computed the mutual information (i.e., the maximum achievable information rate) for an MC channel whereby the information was encoded into the release time of molecules \cite{eckford2007nano}, and by a set of distinct molecules \cite{eckford2007achievable}. In a followup work, Eckford also provided tractable lower and upper bounds on information rate of one-dimensional MC system \cite{eckford2008molecular}.  
In another work \cite{eckford2012molecular}, Kadloor et. al. considered an MC system inside a blood vessel and introduced a drift component into the MC channel to take into account the blood flow, and computed the information rate for the case when pulse-position modulation is used by the emitter. Last but not the least, \cite{srinivas2012molecular} reported an important finding whereby it was proved that the noise in the one-dimensional MC channel with positive drift velocity is additive with inverse Gaussian (IG) distribution. 

Below, we summarize the information rates achieved by some very prominent MC channels.

\begin{itemize}
	\item \textit{Timing Channel:} In a timing channel, the point transmitter encodes a message in the release time of a molecule, and once a molecule reaches the receiver, it is fully absorbed, thus the first arrival time determines the actual arrival time of the molecule. 
	For a single molecule released at time $X$, its actual arrival time $Y$ will be expressed as  \cite{srinivas2012molecular,Li2014CapacityOT}
	\begin{align}
	Y=X+{\mathcal{N}}_T,
	\end{align}
	where ${\mathcal{N}}_T$ is the first arrival time at the receiver boundary.
	For the positive drift $v>0$, ${\mathcal{N}}_T$ follows AIGN distribution IG($
	\frac{l}{v},\frac{{2{l^2}}}{D}
	$) with the communication distance $l$ and diffusion coefficient $D$.
	Based on 
	\begin{align}
	C=\mathop {\max }\limits_{{f_X}\left( x \right):E\left[ X \right] \le m} I\left( {X,Y} \right),
	\end{align}
	\cite{srinivas2012molecular} bounded from above and below the capacity of additive IG noise channel with a constraint on the mean of the transmitted message X.
	Extended from \cite{srinivas2012molecular}, the authors in \cite{Li2014CapacityOT} studied the capacity of the same additive IG noise channel under
	either  an average- and a peak-delay constraint or  a peak-delay constraint,  and the authors in \cite{FarsadMEG16} revisited the capacity bounds of diffusion-based timing channel (without drift) with finite particle's time.
	
	\item \textit{Concentration-encoded Channel:}  In this channel, concentration of molecules is varied to convey information \cite{akan2007an,moore2009molecular,kuran2012inter,pierobon2013capacity}.
	The authors in \cite{akan2007an} studied the mutual information of a more specific molecular communication system with ligand-binding receptors, where the molecules can bind or unbind from the receiver, but without taking into account the diffusion propagation and channel memory. 
	The authors in  \cite{moore2009molecular} modeled and measured the information rate of various molecular communication systems with diffusion, connected, or hybrid-aster propagation approaches, and noise-free, all-noise, exponential decay, and receiver removal noise model. The achievable rates of the diffusion-based MC channel, under two different coding schemes were studied in 
	\cite{kuran2012inter}.
	
	\cite{pierobon2013capacity} considered concentration encoding at the emitter, a diffusion-based MC channel with memory and noise at the receiver, and derived the closed-form expression for the channel capacity. 
	To account for memory, the bounds on capacity of the conventional memoryless Poisson channel was extended to that of the Linear Time Invarient-Poisson channel of diffusion-based single-hop networks \cite{aminian2015ltipoisson}. However, the reception process has not been treated in \cite{pierobon2013capacity,aminian2015ltipoisson}. 
	
	\item \textit{Biological System:} 
	In \cite{eckford2016capacity} and \cite{eino2013design}, the capacities of an inter-cellular signal transduction channel and  bacterial communication were studied by modelling the ligand-reception processes as a discrete-time Markov model, and a Binomial Channel for a bacterial colony, respectively. The capacity analysis of molecular communication channel in a  drug delivery system \cite{molecular2014chahibi} and cell
	metabolism \cite{max2017cell} were studied using COMSOL Multiphysics and KBase (Department
	of Energy Systems Biology Knowledgebase) software application
	suite, respectively. More detailed literature review on information theoretic study of molecular commmunication can be found in \cite{gohari2016information}.
\end{itemize}


\subsubsection{Bit  Error Rate} \label{MC_ber_bep}
 
During each slot, the receiver will receive the molecules due to the current slot as well as from the previous slots (due to brownian motion of molecules). This phenomenon is known as inter-symbol
interference (ISI). As the main bottleneck of bit error performance of   molecular communication system, the ISI is first characterized in \cite{mahfuz2011}, and increasing attention has been focused on the bit error rate performance characterization from then on.

\begin{itemize}
	\item \textit{Single-Hop System with the Passive Receiver:} Initial MC works have focused on a passive (spherical) receiver that just counts the number of received molecules in its close vicinity without interacting with them.
	The bit error rate of the MC system with a passive receiver under ISI and no ISI was studied in \cite{SHAHMOHAMMADIAN2012183} where the receiver implements the optimal maximum a-posteriori probability (MAP) rule. 
	To improve the BER performance of the MC systems, \cite{shih2013channel} introduced a new family of ISI-free coding with fairly low decoding complexity. While, \cite{kilinc2013receiver} did the MAP based, maximum likelihood (ML) based, linear equalizer/minimum mean-square error (MMSE) based, and a decision-feedback equalizer (DFE) based sequence detection. \cite{adam2014receiver} introduced the enzyme reactions to the diffusion, derived the average BER, and verified it via the realistic particle-based simulation. All these works point to the undesirable effect of ISI on the performance of an MC system with a passive receiver.\\
	
	\item \textit{Single-Hop System with the Active Receiver:}
	In a real biological system, the receiver
	actually consists of receptors that react to some specific molecules (e.g., peptides or
	calcium ions). Thus, research efforts have shifted to the simulation and modelling of the active receivers, such as the fully absorbing receiver \cite{yilmaz2014three}, the reversible absorbing receiver \cite{deng2015modeling}, and the ligand-binding receiver \cite{arman2016compre}. 
	\cite{yilmaz2014three} derived a simple expression for the channel impulse response of an
	MC system with an fully absorbing receiver, and validated it by the particle-based simulation simulator (MUCIN).
\cite{deng2015modeling} and \cite{arman2016compre} derived the analytical expressions for the expected received signal and the average BER for an MC system with reversible absorbing receiver,  and for an MC system with the ligand-binding receiver, respectively. The expressions obtained were then verified by particle-based simulation algorithms.\\
	
	\item \textit{Multi-Hop  System and Large-scale System:}
	The average BER of the multi-hop decode-and-forward relay and amplify-and-forward relay MC systems were derived and simulated in \cite{arman2015mult} and \cite{arman2015amplify} to extend the transmission range and improve the reliability of MC systems. Using the three-dimensional stochastic geometry, the average BER with large number of transmitters perfrom joint transmission to the fully absorbing receiver were analyzed and simulated via particle-based simulation and Pseudo-Random simulation in \cite{deng2017analyzing}, which provided an analytical model of BER evaluation for large-scale MC system with all kinds of active receivers.\\
	
     \item \textit{Experimental System:}
      The BER performance of the F$\ddot{o}$rster Resonance Energy Transfer (FRET) nanoscale MIMO communication channel has been tested and examined in \cite{kuscu2015fluorescent}, which was shown to  provide acceptable reliability with BER about 5.7$\times10^{-5}$ bit$^{-1}$ for nanonetworks up to
	150 kbps transmission rates.
	
\end{itemize} 

\subsubsection{Symbol Error Rate}
The symbol error rate (SER) of molecular communication system was first mentioned in \cite{yeh2012frontier}, then the SERs of an MC system with absorbing receiver under the binary concentration keying (BCSK), the quadrature CSK (QCSK), the binary molecular frequency shift keying (BMFSK), and the quadrature MFSK (QMFSK) were simulated in \cite{YILMAZ2014136} using MUCIN simulator. In \cite{chou2015maximum}, the SER of diffusion-based MC system with receiver having periodically ON and OFF receptors and analog filter for computing the logarithm of the
MAP ratio was studied.

\subsubsection{Energy Cost}
\cite{KURAN201086} develops an energy model for the MC system whereby the energy costs in the messenger molecule synthesizing process, the secretory vesicle production process, the secretory vesicle carrying process, and the molecule releasing process were defined based on molecular cell biology. The energy model of vesicle-based active transport MC system was described in \cite{farsad2016energy}, where the energy costs of the vesicle synthesis, the intranode transportation, the DNA hybridization, the vesicle anchoring, loading, unloading, and the micro-tubule motion were defined. In \cite{yansha2017enabling,weisi2017smiet}, a detailed mathematical model for the molecule synthesis cost in MC system with the absorbing receiver was provided to examine the energy efficiency of different relay schemes. In \cite{qiu2017bacterial}, the energy costs in the encoding and synthesizing  plasmid, the plasmid transportation, the carrier bacterial transportation,  the decapsulation and decoding were defined and examined within bacterial relay MC networks.

\section{Enabling and Concomitant Technologies}

\subsection{EM Aspects}
\subsubsection{Nano-Devices}
\begin{figure}[!t]
\centering
\includegraphics[width=0.5\textwidth]{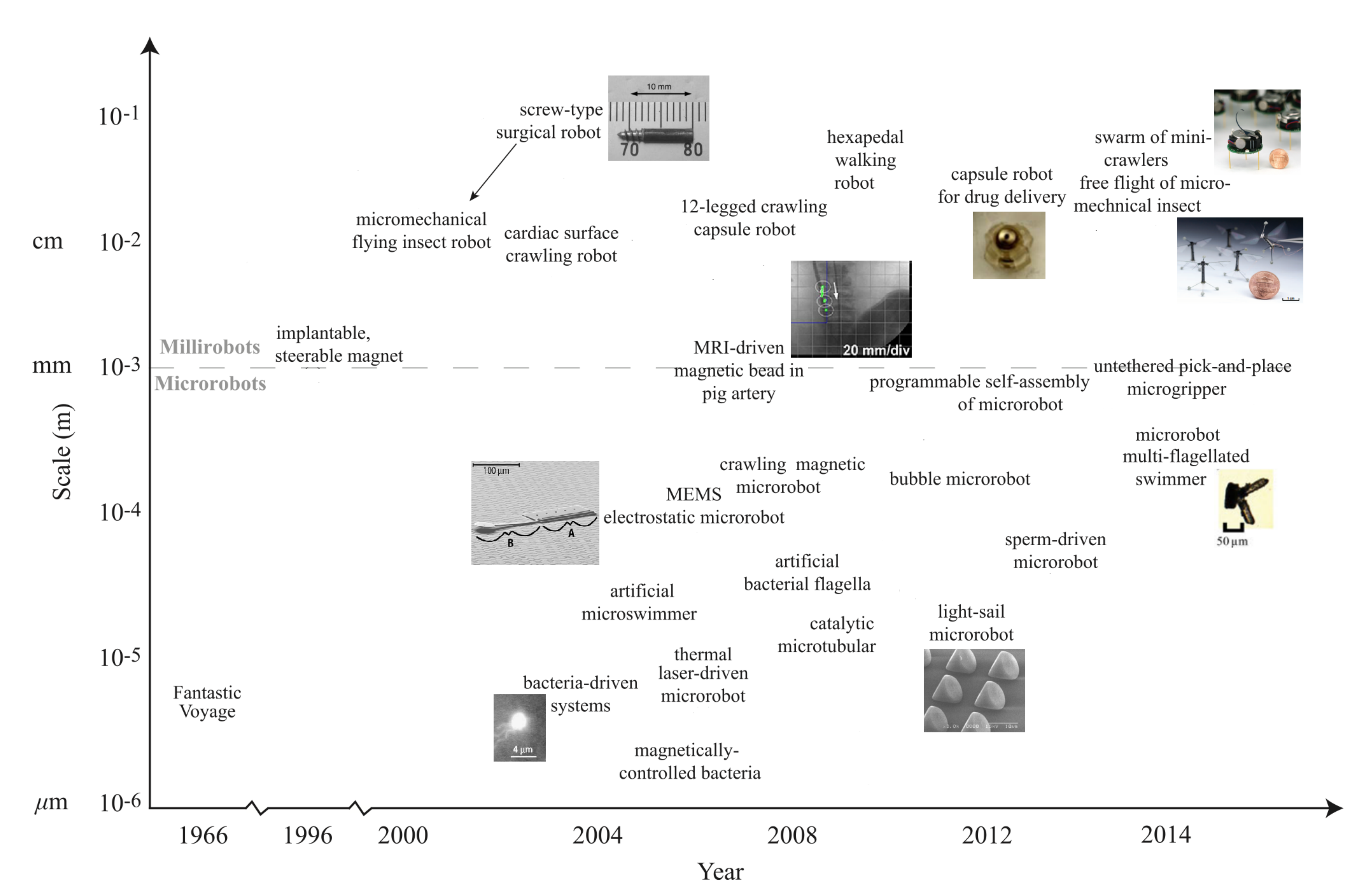}
\caption{Development routine of the micro/nano-devices \cite{kephdthesis}}
\label{fig:develop}
\end{figure}

Advances in nanotechnology have paralleled developments in Internet and sensing technology. The development routine is summarised in Fig. \ref{fig:develop}. At the same time, due to the general belief that graphene/CNT would be the future star of the nano-technology world since its appearance, more attentions has been put on such novel materials and great advances have been achieved. Antenna, as the basic element in communication system, is firstly fully investigated with numerous papers on the design of the antenna made of graphene or CNT in the last five years. First, the possibility of the applications of graphene and CNT was investigated \cite{hanson2005fundamental} and the wave performance on a graphene sheet was also studied in \cite{hanson2008dyadic}. Then, various antennas like graphene patch antenna with different shapes\cite{lin2010100,bala2016development,bala2016investigation}, CNT dipole antenna \cite{hanson2005fundamental,burke2006quantitative,jornet2010graphene-based}, and so on were proposed. Furthermore, a nano-antenna with the shape of log-periodic tooth made of graphene was proposed in \cite{yang2016broad} and a novel graphene-based nano-antenna, which exploits the behaviour of Surface Plasmon Polariton waves in semi-finite sized Graphene Nanoribons (GNRs) was proposed in \cite{jornet2013graphene-based}. Recently, a beam reconfigurable multiple input multiple output (MIMO) antenna system based on graphene nano-patch antenna is proposed in \cite{xu2014design}, whose radiation pattern can be steered dynamically, leading to different channel state matrices. Meanwhile, the design of the sensors made of graphene is also introduced. Reference \cite{kulkarni2014graphene} introduces a graphene-based wearable sensor which can be used to detect airborne chemicals and its concentration level like acetone (an indicator of diabetes) or nitric oxide and oxygen (a bio-marker for high blood pressure, anemia, or lung disease). Later, the sensor made of graphene is designed with higher accuracy to detect HIV-related DNA hybridization at picomolar concentrations, which is a charge detector fabricated of graphene capable of detecting extremely low concentration of charges close to its surface \cite{Fue1701247}. A Stochastic Resonance based (SR-based) electronic device, consisting of single-walled carbon nanotubes (SWNTs) and phosphomolybdic acid (PMo12) molecules, has been developed at Osaka University to apply in bio-inspired sensor \cite{fujii2017single}. It is believed by the authors that by using such devices neural networks capable of spontaneous fluctuation can be developed.

\subsubsection{Internet-of-Things}
Internet-of-Things (IoT) refers to a network of devices with Internet connectivity to communicate directly without human-intervention in order to provide smart services to users \cite{al2015internet}. The Internet-of-Things shares the same development route with nanonetworks, and it is believed that the ultimate goal is to emerge both technologies to form the Internet-of-Nano-Things (IoNT) \cite{balasubramaniam2013realizing}. It is generally believed that the achievements in IoT can also be applied to nanonetworks with minor modification. In IoT, the number of sensors/devices could achieve as high as $tons$ \cite{al2015internet}, many challenges related to addressing and identification of the connected devices would appear, same as nanonetworks. Furthermore, huge amount of data would be produced by such high numbers of sensors which requires high bandwidth and real-time access. Furthermore, implementation of IoT is complex, as it includes cooperation among massive, distributed, autonomous and heterogeneous components at various levels of granularity and abstraction \cite{crooks2012integration}. Applications in health \cite{tyagi2012efficient}, smart security, and smart cities found their way to the market and realize the potential benefits of this technology \cite{keertikumar2015evolution}. In addition, many other applications of IoT can be enumerated such as agriculture, industry, natural resources (water, forests, etc.) monitoring, transport system design, and military applications \cite{khan2012future}.

Network densification is considered as an enabler for the successful diffusion of IoT services and application in the society. In reality, millions of simultaneous connections would be built in IoT, involving a variety of devices, connected homes, smart grids and smart transportation systems \cite{khan2012future}. The concept of cloud and fog computing is introduced to offer large storage, high computation and networking capabilities \cite{nastic2014provisioning}. Also, a high level design of cloud assisted, intelligent, software agent-based IoT architecture is proposed in \cite{crooks2012integration}. Besides of the concept of IoNT, Social Internet of Things (SIoT) is also proposed recently \cite{atzori2012social}. To advocate a common standard, IoT Global Standards (IoT-GSI) are proposed by ITU-T \cite{internet2015url}.


\subsubsection{Bio-Tissue Characterization}
Characterization of channel medium is an essential part to investigate the channel; therefore, it is important to obtain the parameters of bio-tissues if the body-centric communication is under study. Usually, the electromagnetic parameters, \textit{i.e.}, permittivity $\epsilon$ and permiability $\mu$, are used to describe medium in microwave and RF frequency; while at optical frequency, the material is usually described by refractive index (or index of refraction). The techniques such as resonant cavity perturbation method, Transmission-Reflection-Method (TRM), and THz Time Domain Spectroscopy system have been applied to obtain the dielectric property of human tissues \cite{yang2017collagen}. In \cite{Andreuccetti1997internet}, the database of the parameters for human tissues (skin, muscle blood bone and etc.) from 10 Hz to 100 GHz are illustrated, mainly on the basis of Gabriel's work \cite{gabriel1996dielectric,gabriel1996dielectric2,gabriel1996dielectric3}. THZ TDS system is fully studied by E. Pickwell \cite{pickwell2008recent,huang2008tissue,parrott2015terahertz,he2017adaptive} and has been applied to measure the dielectric parameters of bio-tissues like livers \cite{sy2010terahertz}, human colonic tissues \cite{reid2011terahertz}, human breast tissues \cite{ashworth2009terahertz}, \textit{etc.}. Both basal cell carcinoma and normal skin are measured by C.Bao to investigate the possibility of the detection of skin cancer at early stage \cite{truong2015potential} based on the work of parameter extraction of skin with global optimization method \cite{truong2013debye}. And also, the model of human breast tissue in THz band is studied in \cite{truong2015dielectric}. Recently, the performance of DED samples and collagen have been investigated in \cite{chopra2016thz,chopra2016fibroblasts} and the corresponding model has been studied as well to investigate the possibility of adoption of collagen and DED sample as the phantom during the measurement \cite{yang2017collagen}. More work needs to be done to build the database and the appropriate phantom should be sought to use in the measurement setup.

\subsection{Molecular Aspects}
\subsubsection{Molecular Test-beds}
Until now, one fundamental challenge in the application of molecular communication is that we still do not have well studied nano-size biological friendly molecular communication transceivers, 
despite the existing research efforts in  designing and building MC test-beds \cite{farsad2013tabletop,koo2016mimo,krish2013time,leo2013communication,yansha2017ffl,kuscu2015fluorescent,abbasi2018controlled}, and in engineering  biological MC systems \cite{weiss2000engineered,liu2015functionalizing}.
\begin{itemize}
	\item \textit{Macroscale MC Test-beds:} The first macro-scale experimental test-bed for molecular communication was shown in \cite{farsad2013tabletop}, where the text messages were converted to binary sequence, and transmitted via alcohol particles based on a time-slotted on-off-keying modulation. In this tabletop MC test-bed the messages transmission and detection were realized via the alcohol spray and the alcohol metal-oxide sensor, and the message generation and interpretation were electronically controlled via the Arduino micro-controllers. They shown that a transmission data rate of 0.3 bit/s with the bit error rates of 0.01 to 0.03 can be achieved using this single-input-single-output (SISO) tabletop MC test-bed. Later on, this  SISO test-bed  was duplicated to form a multiple-input-multiple-output (MIMO) tabletop MC test-bed with multiple sprays and sensors in \cite{koo2016mimo}, which  achieved 1.7 times higher transmission data rates than that of the SISO test-bed. \\
	
	\item \textit{Nanoscale MC Test-bed:}
	The first nanoscale molecular communication based on the F$\ddot{o}$rster Resonance Energy Transfer (FRET) was implemented and tested in \cite{kuscu2015fluorescent},  where the information was encoded on the energy states of fluorescent molecules, and the energy states were exchanged via FRET.\\

	\item \textit{Microfludic MC Test-beds:} 
	In \cite{krish2013time}, the genetically engineered Escherichia coli (\textit{E. coli}) bacteria population housed in micrometer sized chambers were used as MC transceivers connected via microfluidic pathways, and the message molecule  (N-(3-Oxyhexanoyl)-L-homoserine lactone,
	or C6-HSL) generation and detection were realized via the LuxI enzyme catalyzes and the LuxR receptor protein with fluorescent light based on On-Off Keying (OOK). To improve the achievable data rates of this testbed with OOK, the time-elapse communication (TEC) was proposed by encoding the information in the time interval between
	two consecutive pulses, which shown an order of magnitude data-rate improvement. \\
	In \cite{leo2013communication}, the Hydrodynamic Controlled microfluidic Network (HCN)   fabricated in  poly(dimethylsiloxane) (PDMS) polymer was proposed, where the information was encoded and decoded based  on the distance between consecutive droplets, and droplets carrying information were controlled and transported in HCN to realize molecular communication. The maximum information rate of HCN was analyzed, the noise effect in HCN was simulated using OpenFOAM, and a HCN prototype was fabricated in poly(dimethylsiloxane) (PDMS) polymer.\\
	Inspired by the biological circuits in synthetic biology, a  chemical circuits based on a series of faster chemical reactions were designed to  achieve the transformation of the timing varying information molecules flow from the digital signal to the analog signal inside a designed microfluidic devices in \cite{yansha2017ffl}. This work provides a novel research direction for performing signal processing using chemical circuits inside microfluidic device, and also an alternative method for proof-of-concept analogues of biological circuits with potentially higher speed.\\

\end{itemize}	
	
\subsubsection{Molecular Experiments}
\begin{itemize}
	\item \textit{In Vivo Nervous System Experiment:}
	The first controlled information transfer through an \textit{in vivo} nervous system was demonstrated in \cite{abbasi2018controlled}. 
	Modulated signals were transmitted into nervous systems of earthworms from anterior end, and propagated through earthworms' nerve cord. Although the network of neurons, \textit{i.e.}, the channel response, were considered as a black-box, the authors found the received signals can be decoded as the number of average nerve spikes per input pulse counted in the posterior end. In addition, 
	the MC system was optimized in terms of frequency, amplitude, and modulation scheme, and the authors showed that the data rate can reach $52.6646$ \textit{bps} with a $7.2\times 10^{-4}$ bit error rate when employing a 4FSK modulation and square shaped pulse.\\
 
	  
	\item \textit{Biological MC Experiments:}
	The first engineered intercellular MC  experiment between living bacterial cells  was reported  in \cite{weiss2000engineered}, where the plasmid pSND-1  was the  sender constructed to produce the  autoinducer chemical (VAI)  via the LuxI gene expression inside \textit{E. coli}. Then, the VAI (information messenger) migrates through the cell membranes and medium to interact with the LuxR gene of the receiver-plasmid pRCV-3 inside \textit{E. coli}, and produces Green fluorescent protein (GFP) for information decoding.\\ 
	Using the protein engineering and synthetic biology, a simple MC based on baterial quorum sensing  (QS) was engineered in \cite{liu2015functionalizing}, where a multidomain fusion protein with QS molecular signal generation capability was fabricated as the sender, and an \textit{E. coli} was engineered as the receiver  to receive and report this QS signal. These research demonstrated  the great potential of bio-fabrication of MC devices.
\end{itemize}

\section{Architecture of EM and Molecular Body-Centric nanonetworks}

Generally, it is believed that both EM and MC should share the same network architecture, but will have minor differences according to various specific applications.

\subsubsection{Network Deployment}
\begin{figure}
	\centering
	\includegraphics[width=0.5\textwidth]{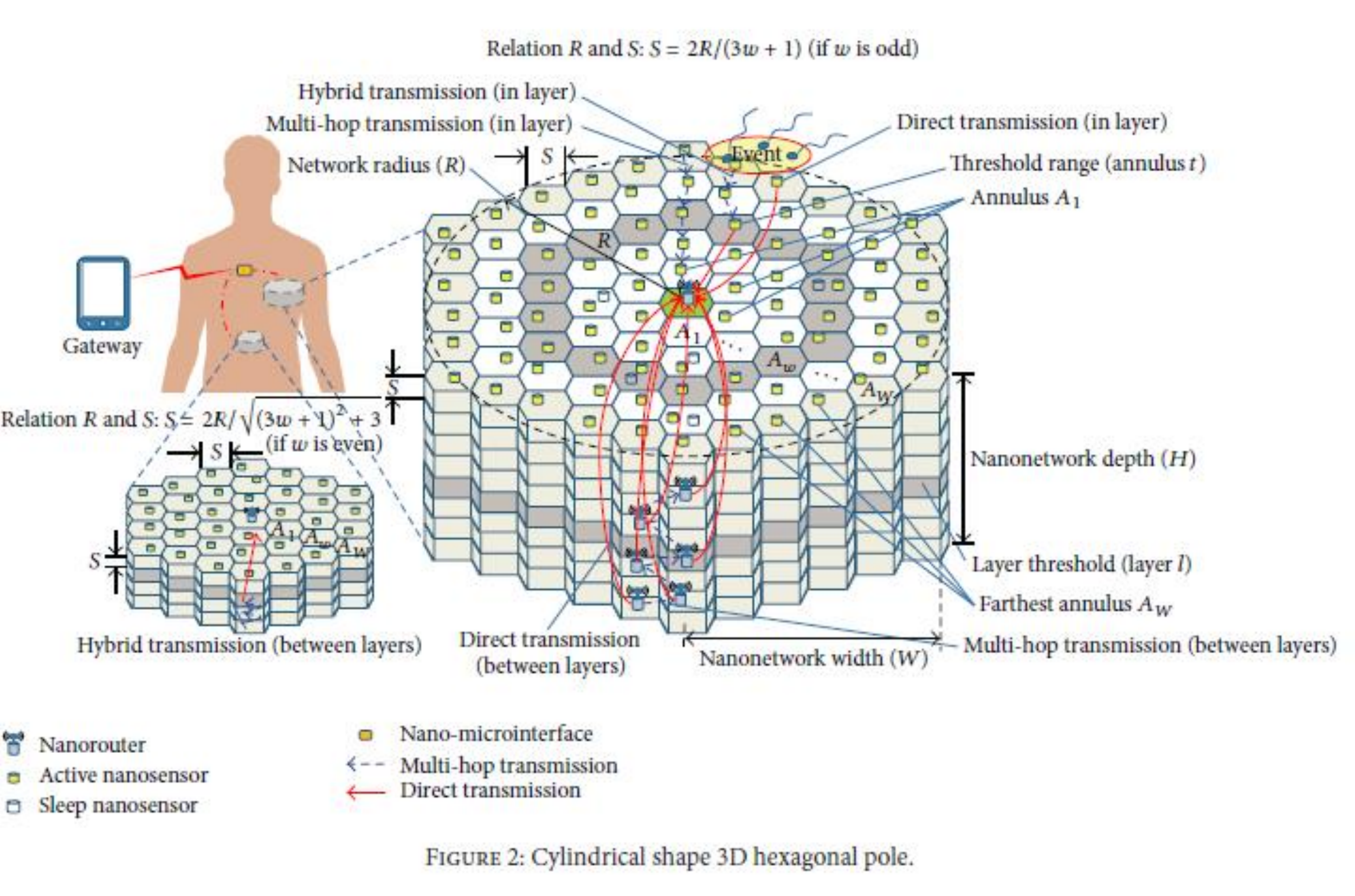}
	\caption{Cylindrical shape 3D hexagonal pole.}
	\label{figure:najah_figure3}
\end{figure}

Aligned with the IEEE P1906.1 framework, the authors of \cite{Jornet} provided an overview of the nanonetworks and is divided in to nano-routers, nano-nodes, gateway and nano-micro interfaces. The work proposed in \cite{Suk} attempts to investigate the ideal number of devices, optimal edge length relative to horizontal length of a general human body organ. The proposed scheme assumes nano-sensors are distributed in a 3-dimensional space in the nanonetworks according to a homogeneous spatial Poisson process as shown in Fig. \ref{figure:najah_figure3}. Authors represent the network deployment as cylindrical 3D hexagonal pole, claiming that the cylindrical shape is closer to the shape of human body organs. They assume that they can put as many nano-sensors as possible and there is only one active nano-sensor in each hexagonal cell. They proposed a scheme for each sensor duty cycle with the assumption that only one sensor is active in each cell. A cell is defined as the smallest living unit of an organ. The ideal number of nano-sensors is calculated using an equation derived by the authors. The equation describes the diameter of the cylinder, the width of the organ in relation to the edge length of the cylinder. The work of \cite{Suk} is considered a step forward in realizing the nano-sensors deployment; however, the authors  assume that all the nano-sensors may recognize other neighbouring nano-sensors. The authors also assume that network deployment also includes routeing nodes; however, they did not state how to calculate the number of routers or micro-interfaces and the positioning technique for these nodes. 

Emre et. al. \cite{ Emre } debates that the first step in network design and deployment is highly tied to the parameters of the nano-antenna, hence nano-antenna design is a critical component of the network design. The reason behind this is their observation that there is a clear trade-off between the number of different tasks the nanonetworks can execute and the reliable communication over the network. Hence, the authors proposed a network of nano-devices that are able to carry out binary tasks and proved that it is possible to construct multi-hop nanonetworks using simple individual nodes activated simultaneously over a shared medium without a significant detriment in reliability. The number of nodes depends on the number of complex tasks for the nanonetworks.  The authors did not provide a mechanism describing the process of choosing the appropriate number of nano-nodes or interfaces. Additionally, the authors did not provide an analysis of the nano-router or interfaces as they did for nano-sensors.  Dressler and Fischer \cite{DRESSLER} discussed the requirements and challenges of designing the gateway or the interface between the nanonetwork and the macro/micro network to bridge the gap of the gateway or interface void. They stated that multiple gateways are required in IoNT deployment such that each one of them is associated with one or more nanonetworks. They also suggested that a gateway should operate at the application layer and recognize the right nanonetwork to receive a message.  They also suggested that the gateway being equipped with one or more nano communication interface should contain the molecular and terahertz interface. As a molecular network may prove to be a significant challenge, a reasonable approach might be to make the gateway an implantable micro device that uses electromagnetic wireless communication to interface the molecular network to the Internet. While the study in \cite{DRESSLER} discussed the requirement and challenges of gateway deployment, they did not provide any solution. Similar to \cite{ DRESSLER }, the study presented in \cite{ Balasubramaniam} discussed the challenges and requirements for the gateway deployment. The study concluded that the gateway will be an implantable device equipped to enable communication with the molecular interface as well as the EM nanonetworks. However, the study remarked that the high ratio of nano-sensors to gateways could lead to swift energy depletion if gateways process information from every nano-sensor. They suggested to thereby distribute the sink architecture and develop a two-layered hierarchy consisting of gateways and nanonetworks.

The aforementioned research attempts to address the network deployment, however, the proposed schemes provide partial solutions to the network deployment; some focused on nano-sensor deployment, while others discussed the requirements and challenges of gateway deployment. However, no all-encompassing solution has been provided in literature yet. Additionally, deployment that achieves essential goals such as survivability, reliability, accuracy or latency intolerance remains an unexplored area of research in nanonetworks deployment.

\subsubsection{Network Mobility}

Nano sensors (NS) are dynamic components in their applications whereby they are forced to move and each move is dictated by their environment. Environmental NS will move according to wind direction and force, which in turns will act to adjust their controller association, location and link quality. Comparatively, the motion of blood monitoring NS will be influenced by its surroundings, whereby speed and turbidity of blood flow and vessel thickness will affect NS link communication quality, velocity and location. This effect is highly pronounced in nanonetworks when compared to traditional sensor networks due to the unique nature of the NS and the used modulation used in nanonetwork communication. Nanonetworks communicate using TS-OOK. This requires nodes to be highly synchronized, an aspect that can be significantly affected by changes in NS mobility. TS-OOK synchronizes transmissions between sender and receiver by requiring the receiver to listen to transmissions on fixed time intervals, thereby ensuring that transmitted bits are received. Distance between the receiver and sender has the largest impact on this process and deciding the time intervals at which the receiver should listen. This distance may change due to NS movement and might result in missing a transmission. The work in \cite{Zainuddin} studied the effect of NS movements on the communication link. The authors studied the pulse time-shift, which is defined by the authors as the distance in time between the actual arrival of the signal and its estimated arrival (in case of no movement), taking into account the Doppler effect, information reduction, and error rate increase. The authors concluded that the doppler effect can be negligible; however, the pulse time-shift can introduce inter-symbol interference (ISI), and the NS movement influences the maximum information rate and the attainable error rate.  The work presented in \cite{Zainuddin} provides a good insight on the effect of mobility in nanosensors networks. However, the assumption of the authors that the transmitter is static while the receiver is mobile, and NSs are moving with the speed of light may limit the scope of the results and their adaptability into applications. 

Even though the mobility of NSs may pause a major challenge on practical deployment of nanosensors, this area is still severely under-researched. References \cite{Petrov} remarked that there is an eminent need to come up with mobility perdition models; a reactive response to NS movement is no longer satisfactory. The authors of \cite{Choi} proposed a movement control method for nanonetworks which is self-organised. The algorithm uses the localization of a particle and its neighbouring paticles to optimise the location of particles and enhance movement positions fo NS through the use of particle swarm optimisation. The proposed algorithm cannot be considered a general mobility model for nanosensors because the algorithm is proposed for homogeneous networks, which is not the norm of a nanonetwork; they are expected to consist of heterogeneous devices with diverse capabilities. Additionally, the model is designed based on the unit disk coverage thereby inheriting the advantages and disadvantages of using this method. In  \cite{Rikhtegar}, the authors proposed a scheme for the hand-off of mobile sensors to the most appropriate nano-controller to conserve energy consumption and reduce the unsuccessful transmission rates. The authors presented a TDMA-based MAC protocol simple fuzzy logic system to control the mobility procedure. They used locally available metrics at each nano-node consisting of the distance of mobile nano-node from nano-controller, traffic load and residual energy of nano-controller, which are considered as fuzzy input variables to control the hand-off decision procedure. The scope of the offered solution is limited by the assumption of constant velocity of the nanosensors and the unit disk transmission similarly to the other proposed schemes. Additionally, the practicality of the system deployment is highly dependent of the trade-off between accuracy and complexity of the algorithm. Hence, the problem of NSs mobility modeling still stands as an urgent area of research for practical deployment of nanonetworkss.

\section{Communication and Networking of EM and Molecular Body-Centric nanonetworks}

\subsection{EM-Based Body-Centric Nanonetworks}

\subsubsection{Physical Layer and MAC Layer} \label{THz_channel_model}
\paragraph{Path Loss Model} 
Studies on THz channel modelling of nano-communication is conducted in \cite{yang2015numerical,abba2016,adva2016}, based on the researches of the one in the air \cite{Chan2010,chan2011,capa2014,femt2014}. From the above studies, it can be concluded that there are three parts in the path loss of the THz wave inside human tissues: the spread path loss $PL_{spr}$, the absorption path loss $PL_{abs}$ and the scattering path loss $PL_{sca}$:
\begin{equation}
PL_{total} [dB] = PL_{spr}(f,r) [dB] + PL_{abs}(f,r) [dB] + PL_{sca}(f,r),
\end{equation}
where $f$ is the frequency while $r$ stands for the path length.

The spread path loss, caused by the expansion of the wave in the medium, is defined as
\begin{equation}
PL_{spr}(f,r) = \left( \frac{4 \pi r}{\lambda_g} \right)^2=(\frac{4\pi n_r f r}{c})^2,
\end{equation}
where $\lambda_g = \lambda_o/n_r$ represents the wavelength in medium with free-space wavelength $\lambda_o$, and $r$ stands for the transmission distance of the wave. Generally, the electromagnetic power is considered to travel spherically. $4 \pi r^2$ denotes the isotropic expansion term and $4\pi(\dfrac{n_rf}{c})^2$ is the frequency dependent receiver antenna aperture term.

The absorption path loss represents the attenuation absorbed by the molecular of the medium. It is assumed that part of the energy would convert into internal kinetic energy to excite the molecules in the medium. By reversing the the transmittance of the medium $\tau (f,d)$, we can obtain the absorption loss:
\begin{equation}
\label{eq:PL_abs}
PL_{abs} = \frac{1}{\tau(f,r)} = e^{\alpha(f) r},
\end{equation}
where $\alpha$ is the absorption coefficient while $r$ is the distance.

The scattering path loss accounts for the loss of the signals caused by the deflection of the beam because of the non-uniformity's of the environment. Take human as an example, there are tons of molecules, cells, organs with various shapes and EM properties. The effects are dependent not only on the size, shape and EM properties of the particles but also on the wavelength of the transmitting signal. In \cite{Elayan2017Terahertz}, the detailed phenomenon was discussed and it can be written as
\begin{equation}
PL_{sca}(f,r) = e^{- \mu_{sca} r},
\end{equation}
where $\mu_{sca}$ refers to the scattering coefficient and $r$ is the travelling distance.

In \cite{Elayan2017Terahertz}, the effects of all three path loss have been fully discussed for the in-body nano-communication. It is stated that the scattering path loss is almost negligible compared with the absorption path loss at the THz band.

\paragraph{Noise model}
The molecular absorption noise is the main element of the noises at Terahertz band, which is introduced by the molecular vibration, partially re-radiated the energy absorbed from the EM waves \cite{chan2011}. Therefore, such noises are dependent on the transmitted signal. In \cite{Ener2013}, noise model was investigated while in \cite{zhang2017analytical} noise of the human tissues was studied.

The total molecular absorption noise p.s.d. $S_N$ can considered as the summation of the atmospheric noise $S_{N0}$, the self-induced noise $S_{N1}$ and others originating from other sources like devices $S_{No}$:
\begin{equation}
S_N(r,f)=S_{N0}(r,f)+S_{N1}(r,f)+S_{No},
\end{equation}
\begin{equation}\label{eq:atm_noise}
S_{N0}(f)= \lim_{r\to\infty} k_B T_0 (1-e^{-\alpha(f)r}) (\frac{c}{\sqrt{4\pi}f_0})^2, 
\end{equation}
\begin{equation}\label{eq:self_noise}
S_{N1}(r,f)=\ S(f)(1-e^{-\alpha(f)r})(\frac{c}{4{\pi}fr})^2, 
\end{equation}
where $r$ refers to the propagation distance, $f$ stands for the frequency of the EM wave, $k_B$ is the Boltzmann constant, $T_0$ is the reference temperature of the medium, $\alpha(f)$ is the absorption coefficient, $c$ is the speed of light in vacuum, $f_0$ is the design centre frequency, and $S$ is the p.s.d of the transmitted signal.

The atmosphere can be seen as an effective black body radiatior in the homogeneously absorbing medium; thus, the absorbing atmosphere with any temperature would produce the atmospheric noise\cite{kokkoniemi2016discussion}. Such atmospheric noise is called as the background noise, independent on the transmitted signal. However, the noise model of Eq. \eqref{eq:atm_noise} only describes a special case for THz wave in air. Without loss of the generality, the term $k_B T_0$ should be replaced with the Planck's law, which describes the general radiation of the black body \cite{kokkoniemi2016discussion}. Therefore, the molecular absorption noise contains three main contributors: the background noise $S_{Nb}(r,f)$, the self-induced noise $S_{Ns}(r,f)$ and other noise $S_{No}(r,f)$:
\begin{equation}
S_N(r,f)=S_{Nb}(r,f)+S_{Ns}(r,f)+S_{No}(r,f).
\end{equation}

The detailed discussions were conducted in \cite{zhang2017analytical}, and it is found that the molecular absorption noise would be the essential part of the contributors to the noise at the receiver. Meanwhile, p.s.d of human tissues on the sef-induced noise and background noise are investigated as well, where the following trends were observed:
\begin{itemize}
\item The background noise p.s.d stay steady for all three tissue types because of the slight difference of refractive index.
\item The induced noise p.s.d change slowly with frequency, different from the fierce fluctuations of THz communication in air \cite{chan2011}.
\item The self-induced noise p.s.d is way bigger than the background noise for all three human tissues, leading to the conclusion that the background noise could be neglected in vivo. 
\end{itemize}

\paragraph{Modulation Technique}
Because the limitation of the size, nano-devices are power-limited; thus,  it is not possible to adopt the traditional modulation techniques which would cause energy. Based on such situations, the modulation of carrier-less pulse based modulation is investigated in \cite{perf2015}. And a pulse modulation technique, , named TS-OOK, is studied in \cite{low2011} and improved in \cite{jornet2014femtosecond} to fully exploit the potential of the nano-devices made of graphene. So far, TS-OOK is the most promising communication scheme for resource-constrained nanonetworks.

To investigate the collision between symbols in body-centric nano-communication, reference \cite{zhang2017analytical} investigated the feasibility of TS-OOK as a communication scheme at THz band for the in-body communication of nanonetwork where not only the noise but also the interference is investigated. It shows that the received signal power is closely related to the transmitted signal power; thus we need to choose the transmitted power carefully to make the difference of the received power with the silence pulse large enough to make the detection accurate. In \cite{jornet2011information}, TS-OOK is introduced and femto-second pulse is used as the communication signal between nano-devices \cite{akyildiz2010electromagnetic}. Reference \cite{jornet2014femtosecond} analysed this pulse-based modulation where the transmitted pulse length is 100 $fs$. Meanwhile, the channel access scheme of nano networks at THz band was proposed and analyzed. In the paper, interference-free scenario and multi-user scenario were both discussed. In the end, the model was evaluated by COMSOL Multi-physics \cite{COMSOL}. The results showed that such modulation schemes were suitable for nanonetworks and by choosing suitable parameters the rates would go from a few Gbps to a few Tbps. Later, Rate Division Time-Spread On-Off Keying (RD TS-OOK) is studied in \cite{jornet2012phlame} and the PHysical Layer Aware MAC protocol for Electromagnetic nanonetworks in the Terahertz Band (PHLAME) is first proposed. The proposal of these two concept is to support the extremely high density of nano-devices in nanonetworks and enable the network throughput to go up to tens of Gbps. In 2013, the critical packet transmission ratio (CTR) was derived in \cite{wang2013energy} to introduce an energy and spectrum aware MAC protocol which can make nano-sensors transmit with high speed with little energy consumption.

\paragraph{Coding Technique}

Due to the simple structure, nano-nodes only have limited power storage. Thus, to save the transmitted energy, numerous coding methods were discussed. Fixed-length codewords with a constant weight can be used not only reduce the power consumption, but also to reduce the interference  \cite{jornet2011low}. Kocaoglu \textit{et al.} \cite{kocaoglu2012minimum,kocaoglu2013minimum} proposed a fixed-length coding methods later to keep the  Hamming distance of codewords which would make the Average Code Weight (ACW) lowest. The performance study of the fixed-length code at the aspects on ACW and code length was conducted in \cite{chi2013energy}. Based on this research, variable-length low weight codes for OOK modulation was investigated in \cite{chi2014energy} which would lower the transmission energy while keep the desired throughput.

\subsubsection{Network Layer}
\begin{itemize}
\item \textbf{Addressing} The IEEE 1906 standard defines the specificity as the technique that enables a reception of a message carrier by a target and maps it to an address in classical communication systems. However, it does not provide any discussion on how to generate, manage, or assign specificity component to nanonodes in molecular or EM nanonetworks. Individualized network addresses and conventional addressing are not feasible nor practical due to the nano-scale of the nanonetworks. Therefore, the use of cluster-based addressing is advantageous over node-base addressing. It provides the ability to address a group of nodes with a specific function in monitoring health or in a specific biological organ \cite{Agoulmine}.  Additionally, addressing may be safely assumed to be required in inbound direction within the nanonetworks to inform a cluster or a nanonetwork performing a specific function (application) on its next action. However, in outbound direction, no addressing is necessary since the outbound device is the sink of communication of the nanonetworks; whenever a gateway receives a message from inside, it will simply forward it to that device \cite{DRESSLER}. Hence, conventional addressing is not necessary for nanonetworks. It may be sufficient to reach a destination just to know the right direction, since it may be the only possible option as discussed above or any member of cluster in that direction is a suitable destination. 
Addressing in its conventional meaning may not be needed. For example, broadcasting a message in nanonetworks may be a solution for data dissemination because of the low possibility of collision in THz band due to the wide bandwidth and transmission time. A receiver overhearing the message decides if the message is of interest. This method can be naturally implemented in molecular nanonetworks. Direct connectivity between nanodevices is another example, where a guided communication can be provided via antenna aperture, resonant frequency, or impedance match in EM nanonetworks and shape or affinity of molecule to a particular target, complementary DNA for hybridization, etc in molecular networks. 
In literature, several authors in the context of proposing routing or MAC protocols for EM nanonetworks assumed that the nanonodes are assigned addresses without discussing how (for e.g \cite{FERRANTI, Piro}). Few studies discuss nanonetwork addressing. Stelzner et.al.  \cite{STELZNER} proposed an addressing scheme that is based on the function of the nanosensor and its location rather than focussing on individual nodes.  The authors proposed employing known protocols like IPv6 or overhead-reduced variants like 6LoWPAN for the control station and gateways. In the proposed scheme, it is irrelevant which specific sensor detects an event or which node executes a service as long as the right function is performed at the right location. However, this scheme is challenged when exact and specific quantities are required such as the case in releasing a certain amount of a drug. Addressing a partial number of nodes based on the required quantity with the lack of individual addressing of node continues to be an open area of research.   

\item \textbf{Routing} One of the most fundamental concerns for Body-Centric nanonetworkss is accurate routing in order to transmit signal promptly and precisely. Some challenges affect the routing protocol, including energy, complexity, latency and throughput. Thinking of the limited resource-equipped nano-sensors, one of the most important requirement is to reduce the energy consumption. There have been a few attempts towards achieving energy efficiency in such networks by multi-hop networking \cite{pierobon2014routing,yu2015forwarding,tsioliaridou2015corona}.

A routing framework for WNSNs is proposed to guarantee the perpetual operation while increase the overall network throughput \cite{pierobon2014routing}. The framework uses a hierarchical cluster-based architecture. The choice between direct and multi-hop transmission is determined based on the probability of energy savings through the transmission process. It is concluded that multi-hop performs better for varying distance. However, only two hop counts are considered which requires more hops consideration in the performance evaluation. Besides, it mainly focuses on WNSNs and does not solve the requirements and constraints of BCNN. The primary task of networking protocol is forwarding, which is sending packets to the next-hop along its path to the destination. In traditional wireless sensor networks (WSN), multi-hop forwarding schemes including the nearest hop forwarding, the longest hop forwarding and the random forwarding schemes as well as the single-hop end-to-end transmission are utilised.  For long range THz wireless nano-sensor networks (WNSN) with absorption-defined windows, in order to overcome the frequency-selective feature, a channel-aware forwarding scheme is proposed in\cite{yu2015forwarding}. The selection of the next hop is a trade-off between minmising the transmission distance and the hop count. Nevertheless, all the relay nodes are assumed to have sufficient energy and computation capacity which is impractical. Moreover, authors in \cite{tsioliaridou2015corona,tsioliaridou2016lightweight} propose a geographic routing protocol. User-selected nodes are used as anchor-points at the setup phase, and all nodes measure their distances from these anchors to obtain address. The routing then employs the appropriate subset of anchors which is selected by the sender of a packet. However, the proposed scheme is based on fixed topology neglecting the mobility and dynamic of nano-nodes. A flood-based data dissemination scheme is introduced in \cite{liaskos2015promise}. This scheme classifies each node as infrastructure or network user after processing the reception quality. Only infrastructure nodes can act as re-transmitters, while the remaining nodes revert to receiving-only mode. This approach improves the energy efficiency by avoiding the unconditional broadcast and reliving the serious redundancy and collisions. Nonetheless, this dynamically-forming infrastructure requires topology-dependent optimisation and digital signal processing capabilities of nano-nodes.

BCNN routing protocols design provides a challenge with no real solutions despite the growing research tackling this area. Two kinds of energy-harvesting protocol stacks that regulate the communication among nano-devices are proposed in \cite{piro2015design}. The greedy energy-harvesting scheme simply delivers the packet to the node with the higher energy level, while the optimal energy-harvesting scheme selects the node that can maximise the overall energy level within each cluster. Both schemes shown better performance compared with the traditional flooding-based scheme. However, the optimal routing strategy cannot be easily employed because of its high computational capacity requirement. Besides, the transmission distance is not taken into consideration, which makes the selection of relay path inappropriate only based on the energy level. Recently, a cognitive routing named enhanced energy-efficient approach is proposed for IoNT \cite{al2017cognitive}. An analytic-hierarchy process is implemented as the reasoning element to make the cognitive decision based on observing the dynamically changing topology of the network. 
\end{itemize}

\subsubsection{Transport Layer} 
IEEE P1906.1 standard in mapping the nanonetwork to the conventional layering system ignored the transport layer as shown in Table II. Reliable transmission is a requirement for practical implementation of nanonetworks. Due to the peculiar characteristics of the nanonetworks, researchers agree that reliability at the MAC layer or the transport layer is sufficient but not necessary in both. Hence, the IEEE P1906.1 framework assumes the existence of the MAC layer and the absence of the transport layer. Piro et. al. {\cite{Piro}} implemented two types of MAC protocols, transparent MAC and Smart MAC, in designing their nano simulator. Transparent MAC pushes packets from the network layer to the physical interface without any processing at the MAC layer. Smart MAC enqueues a packet in reception to discover the neighboring nodes before sending the packet through a handshaking procedure. For transparent MAC, researches assume that the reliability service is shifted to the transport layer, thereby advocating for the existence of transport layer. The authors of \cite{Jarmakiewicz} proposed adapting the Optimized Exchange Protocol (OEP) protocol, which is part of the IEEE 11073-20601 standard {\cite{IEEE_11073}} and is particularly important in telemedicine to provide access points to services from the application layer to the transport layer. The OEP protocol is flexible and lightweight, which makes it suitable for implementation in constraint processing power and storage nano devices. However, the authors did not propose any technique on how to adapt or implement the OEP protocol for nanonetworks. Tairin et. al. \cite{Tairin} proposed an acknowledgement-based UDP protocol to improve the packet delivery ratio in a nanonetwork.  The proposed protocol utilizes timeout timer in UDP to double check whether the packet gets delivered to the destination. The authors evaluated the performance of the protocol via simulation and found that the proposed protocol improved the delivery ratio of packets but introduced additional delay to the network.  Few proposals addressed transport layer protocol design. This area remains unexplored in academia and industrial research. The interaction of congestion avoidance and reliability between the MAC layer and the transport layer along with the trade-off of induced delay and energy consumption is yet to be explored.

\subsection{MC-Based Body-Centric Nanonetworks}
\subsubsection{Propagation Channel Model} 
\begin{itemize}
	\item  In the \textit{\textbf{free-diffusion}} channel, the  information molecules (such as
	hormones, pheromones, DNA) move in the
	fluid medium via Brownian motion. In this case, the propagation
	is often assumed to follow the  Wiener process, and the propagation model can be mathematically described using Fick's second law \cite{howard1993radom630}:
	\begin{align}
	\frac{{\partial C}}{{\partial t}} = D{\nabla ^2}C, \label{ficklaw}
	\end{align}
	where  the diffusion coefficient $D$
is governed by the Einstein relation as \cite{philip2008biological587}
	\begin{align}
	 D = \frac{{{k_B}T}}{{6\pi \eta {r_m}}}, \label{D}
	\end{align}
	where $T$ is temperature in kelvin, $\eta$ is the viscosity of the fluid environment, $r_m$ is the radius of information molecule, and $k_B$ is the Boltmann constant. This Einstein relation may lose accuracy in most realistic scenarios, and the diffusion coefficient of which 
	is usually obtained via experiment \cite{philip2008biological587}.\\
	\item  In the \textit{\textbf{diffusion with drift}} channel, the propagation model in a 3D environment can be mathematically expressed as  \cite[Ch. 4]{howard1993radom630}
	\begin{align}
	\frac{{\partial C}}{{\partial t}} = D{\nabla ^2}C - {v_x}\frac{{\partial C}}{{\partial x}} - {v_y}\frac{{\partial C}}{{\partial y}} - {v_z}\frac{{\partial C}}{{\partial z}}, \label{ficklawdrift}
	\end{align}
	where ${v_x}$, ${v_y}$, and ${v_z}$ are the constant drift velocities in the $+x$, $+y$, and $+z$ directions, respectively.\\
	Different from the EM wave propagation model, the molecular propagation has the advantages of not sufferring from the diffraction loss under the shadow of objects, and  not restricting by the cut-off frequency in pipe, aperture, and mesh environments \cite{guo2016molecular}.\\
\end{itemize}	

\subsubsection{Noise Model} 
\begin{itemize}	
	\item 
	The \textit{\textbf{inherent noise}} is usually contributed by the random arrival of emitted molecules at the previous bit intervals.\\ 
	In the \textit{timing} channel, the noise  ${\mathcal{N}}_T$ is the first arrival time at the receiver boundary given as \cite{srinivas2012molecular} 	
	\begin{align}
	{\mathcal{N}}_T \sim \text{IG}(\frac{l}{v},\frac{{2{l^2}}}{D}	),
	\end{align}
	with the communication distance $l$ and diffusion coefficient $D$ for the positive drift $v>0$.
	
	In the \textit{\textbf{concentration-encoded}} channel, the number of left over molecules belonging
	to the previous bit to the current bit duration follows the binomial distributions, and the noise at the $n_b$th bit interval due to previous  ${({n_b} - 1)}$ bit intervals  is described as \cite{KURAN201086,deng2015modeling}
	 \begin{align}
	 {\mathcal{N}}_{n_b} \sim \sum\limits_{i = 1}^{{n_b} - 1} {Binomial\left( {N,F\left( {d,\left( {{n_b} - i} \right){T_b},\left( {{n_b} - i + 1} \right){T_b}} \right)} \right)} ,
	 \end{align}
	 where $N$ is the number of transmit molecules at the start of the first bit interval, $n_b$ is the number bit intervals,  $T_b$ is the length of one bit interval,  $d$ is the distance between the transmitter and the receiver, and  ${F\left( { \cdot , \cdot, \cdot } \right)}$ is the fraction number of molecules counted at the receiver. \\
	 \item The \textit{\textbf{external noise}} usually includes the biochemical noise, the thermal noise, the physical noise,  the sampling noise, and the counting noise.
	 The biochemical noise is the biochemically interaction between the information molecules/bio-nanomachines and the surrouding molecules and environment. The thermal noise is the varied activity levels of the thermally activated processes or stochastic thermal motion due to the changing surrounding temperature, and the physical noise is the physical force on the  molecules movement due to the viscosity of fluid environment \cite{nakano2014mole}.   The  counting noise arises when measuring the molecular concentration at the receiver location, and it is due to 
	 the randomness in the molecules movement and to the discreteness of the molecules, whereas the sampling noise arises when modulating the molecular concentration at the emission of molecules, and is due to the discreteness of the molecules and the unwanted perturbation at the emission process \cite{pierobon2011noise}.\\
	 
	\end{itemize}	

\subsubsection{Modulation Techniques} 

Different from the modulation in radio frequency (RF) wireless communication systems where the information is modulated on the amplitude, frequency, and phase of the radio waves, the molecular communication transmitters modulate the information on the type/structure, the emitting time, and the number of releasing molecules. 
	 \begin{itemize}	
	 	\item  In the \textit{\textbf{timing channel}}, the information was modulated on the emitting time of molecules as in \cite{eckford2008molecular, eckford2012molecular,srinivas2012molecular,Li2014CapacityOT}.\\	 
	 \item
	 	In the  \textit{\textbf{concentration-encoded}} channel, two types of modulation schemes for binary  MC system was first described in \cite{mahfuz2010nano}, which are the ON-OFF modulation and the Multilevel amplitude modulation (M-AM). In the ON-OFF modulation scheme, the concentration of information molecules during the bit interval is $Q$ to represent  bit-1, and the concentration of information molecules during the bit interval is 0 to represent  bit-0. In the M-AM scheme, the concentration of information molecules is continuous sinusoidal wave, where the amplitude and the frequency can be encoded. The Concentration shift keying (CSK) was proposed for modulating the number of information molecules,  and Molecule Shift
	 Keying (MoSK) was proposed for modulating on different types of information molecules \cite{kuran2012inter}.\\
	  Due to the constraints in the accurate time arrival of molecules in random walks, and 
	 the limited types of molecules in MC system, the Binary CSK modulation based on the number of releasing molecules have been 
	  widely applied  \cite{deng2017analyzing,YILMAZ2014136,kuran2012inter,yilmaz2014three,noel2014improving,kim2013novel,deng2015modeling}, where the molecules concentration is considered as the signal amplitude. In more detail, in the  Binary CSK,  the transmitter emits $N_1$ molecules at the start of the bit interval
	 to represent the bit-1 transmission, and emits $N_2$ molecules at the start of the bit interval to represent the bit-0 transmission. In most works, $N_1$ can be set as zero to reduce the energy consumption and make the received signal more distinguishable. The hybrid modulation based on the number as well as the types of releasing moleules were proposed and studied in \cite{arjmandi2013diffusion,tepekule2015isi}.
	\end{itemize}	

\subsubsection{Reception Model}

For the same single point transmitter located at ${\overrightarrow{r}}$ relative to the center of a  receiver with radius $r_r$,
the received number of molecules will be different depending on the types of receivers.\\ 

\begin{itemize} 	 	
\item	For the	\textit{\textbf{passive receiver}}, the local point concentration at the center of the passive receiver at time $t$ due to a single  pulse emission by the transmitter occurring at $t=0$ is given as \cite[Eq. (4.28)]{nelson2004biological}
		\begin{align}
		C\left( {\left. {{ {{\Omega}} _{{r_r}} },t} \right|{\overrightarrow{r}}} \right) = \frac{1}{{{{\left( {4\pi Dt} \right)}^{{3 \mathord{\left/
								{\vphantom {3 2}} \right.
								\kern-\nulldelimiterspace} 2}}}}}\exp \Big( { - \frac{{{{ \left| {\overrightarrow{r}} \right|
						}^2}}}{{4Dt}}} \Big), \label{Fraction_d_pass}
		\end{align}
		where $\overrightarrow{r}  = [x,y,z]$,  and $[x,y,z]$ are  the coordinates along the three axes.

\item	For the	\textit{\textbf{fully absorbing receiver}} with spherical symmetry, the reception process can be  described as{ \cite[Eq. (3.64)]{schulten2000lectures}}
	\begin{align}
	{\left. {D\frac{{\partial \left( {C^{\text{FA}}\left( {r,\left. t \right|{r_0}} \right)} \right)}}{{\partial r}}} \right|_{r = r_r^ + }} = {k}C^{\text{FA}}\left( {{r_r},\left. t \right|{r_0}} \right),  k \to \infty \label{boundary1}
	\end{align}	
	where $k$ is the absorption rate (in length$\times $time$^{-1}$).
The molecule distribution function of the fully absorbing receiver at time $t$ due to a single  pulse emission by the transmitter occurring at $t=0$ is presented as 
	\begin{align}
	C^{\text{FA}}\left( {\left. {r,t} \right|{r_0}} \right) = \frac{1}{{4\pi r{r_0}}}\frac{1}{{\sqrt {4\pi Dt} }}\left( {{e^{ - \frac{{{{\left( {r - {r_0}} \right)}^2}}}{{4Dt}}}} - {e^{ - \frac{{{{\left( {r + {r_0} - 2{r_r}} \right)}^2}}}{{4Dt}}}}} \right),
	\label{der22}
	\end{align}  
	
\item	For the \textit{\textbf{reversible adsorption receiver}} with spherical symmetry,
	the  boundary condition of the information molecules at its surface is  \cite[Eq. (4)]{andrews2009accurate}
	\begin{align}
	{\left. {D\frac{{\partial \left( {C\left( {r,\left. t \right|{r_0}} \right)} \right)}}{{\partial r}}} \right|_{r = r_r^ + }} = {k_1}C\left( {{r_r},\left. t \right|{r_0}} \right) - {k_{ - 1}}{C_a}\left( {\left. t \right|{r_0}} \right), \label{boundary1}
	\end{align}	
	where  $k_1$ is the adsorption rate  (length$\times $time$^{-1}$) and $k_{-1}$ is the desorption rate  (time$^{-1}$), and its molecule distribution function was derived in \cite[Eq. (8)]{Yansha16rever}.  
	
\item	For the \textit{\textbf{ligand binding receiver}} with spherical symmetry, the  boundary condition of the information molecules at its surface is
			\begin{align}
		4\pi r_r^2D{\left. {\frac{{\partial C\left( {\left. {r,t} \right|{r_0}} \right)}}{{\partial r}}} \right|_{r = {r_r}}} = {k_f}C\left( {\left. {r,t} \right|{r_0}} \right) - {k_b}\left[ {1 - S\left( {\left. t \right|{r_0}} \right)} \right],
		\end{align}
		where $S\left( {\left. t \right|{r_0}} \right)$ is the probability that the information molecules released at distance $r_0$ given as
		\begin{align}
		S\left( {\left. t \right|{r_0}} \right) = 1 - {\int_0^t {4\pi r_r^2D\left. {\frac{{\partial C\left( {\left. {r,\tau } \right|{r_0}} \right)}}{{\partial r}}} \right|} _{r = {r_r}}}d\tau,
		\end{align}
		 and its molecule distribution function was derived in \cite[Eq. (23)]{arman2016comprehensive}. 	
\end{itemize}

\subsubsection{Coding Techniques}
Similar to traditional wireless communication systems, many coding schemes have been studied for molecular paradigm to improve transmission reliability. 
Hamming codes were used as the error control coding (ECC) for DMC in \cite{leeson2012forward}, where the coding gain can achieve 1.7dB with transmission distance being $1\mu$m. Meanwhile, the authors modelled the energy consumption of coding and decoding to show that the proposed coding scheme is energy inefficient at shorter transmission distances. In their subsequent work, the minimum energy codes (MECs) were investigated and outperformed the Hamming codes in bit error rate and energy consumption \cite{bai2014minimum}. Moreover, the authors of \cite{lu2015comparison} compared and evaluated the Hamming codes, Euclidean geometry low density parity check (EG-LDPC) and cyclic Reed-Muller (C-RM) codes. In order to mitigate the inter-symbol-interference (ISI) caused by the overlap of two consecutive symbols for DMC, Reed Solomon (RS) codes 
were investigated in \cite{dissanayake2017reed}. Compared with the Hamming codes capable of correcting one bit error, the RS codes are highly effective against burst and random errors. The results showed that the bit error probability (BEP) increases as either the number of molecules per bit increases or the codeword minimum distance decreases.

Besides these frequently used wireless communication codes,
new coding schemes were developed to tailor MC channel characteristics, such as the coding based on molecular coding (MoCo) distance function 
\cite{ko2012new} and the ISI-free code for DMC channels with a drift \cite{shih2013channel}. 
Further to these, the authors of \cite{marcone2017parity} considered coding implementation and designed a parity check analog decoder using biological components. The decoding process depends on the computation of a-posteriori log-likelihood ratio involving L-value and box-plus calculation. The calculations are completed with the help of chemical reactions 
and the gene regulation mechanism whose input-output relation can be described by Hill function. Through carefully choosing the parameters in Hill function,
the Hill function is able to approximate some mathematical operations, such as the hyperbolic operation and logarithmic operation, and finally leads to the successfully bits decoding. More details on coding schemes for MC could refer to \cite{farsad2016comprehensive,kuscu2019transmitter}.

\section{Interconnectivity of EM and Molecular Body-Centric nanonetworks}
\label{hy_com}

\subsection{Requirements and Opportunities}
\label{requirements_opportunities}
Besides the five components discussed in Sec. \ref{standard}, the IEEE P1906.1 framework also defined the element of the interface between the In-Body Network and the Body-Area Network which is an important part for the application implementation of nanonetworks, especially for medical-related applications. However, as the goal of the standard is to highlight the minimum required components and their corresponding functions necessary to deploy a nanonetwork, which communication paradigm is adopted inside the human body and outside people, and what is the interface to transmit healthy parameters from nano-nodes inside human body to outside devices are not specified. 

\begin{figure}[t]
	\centering
	\begin{subfigure}{0.5\textwidth}
		\centering
		\includegraphics[width=0.65\textwidth]{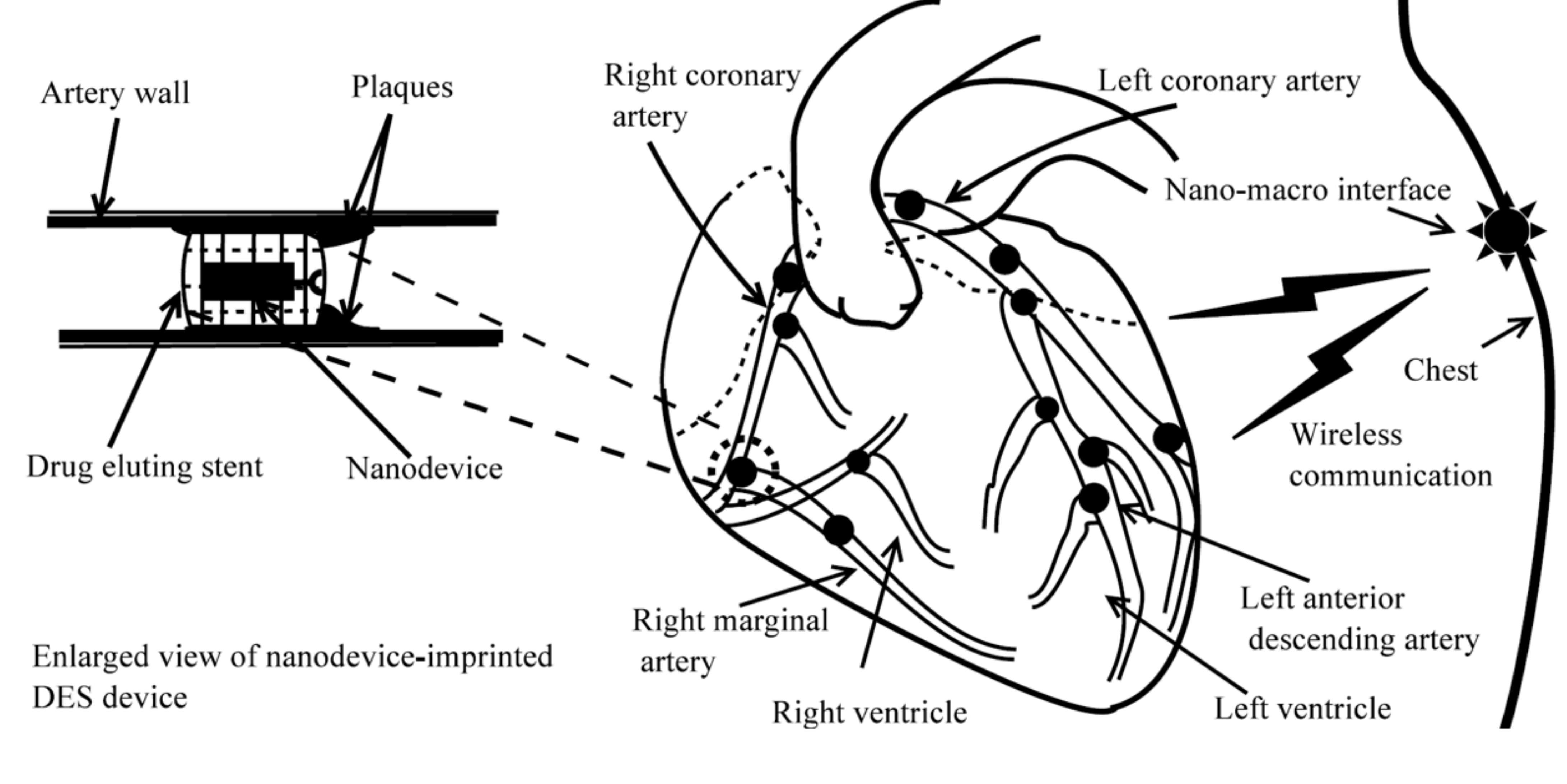}
		\caption{The proposed scheme in \cite{Islam}}
		\label{f_EM_1}
	\end{subfigure} \\
	\begin{subfigure}{0.45\textwidth}
		\centering
		\includegraphics[width=0.85\textwidth]{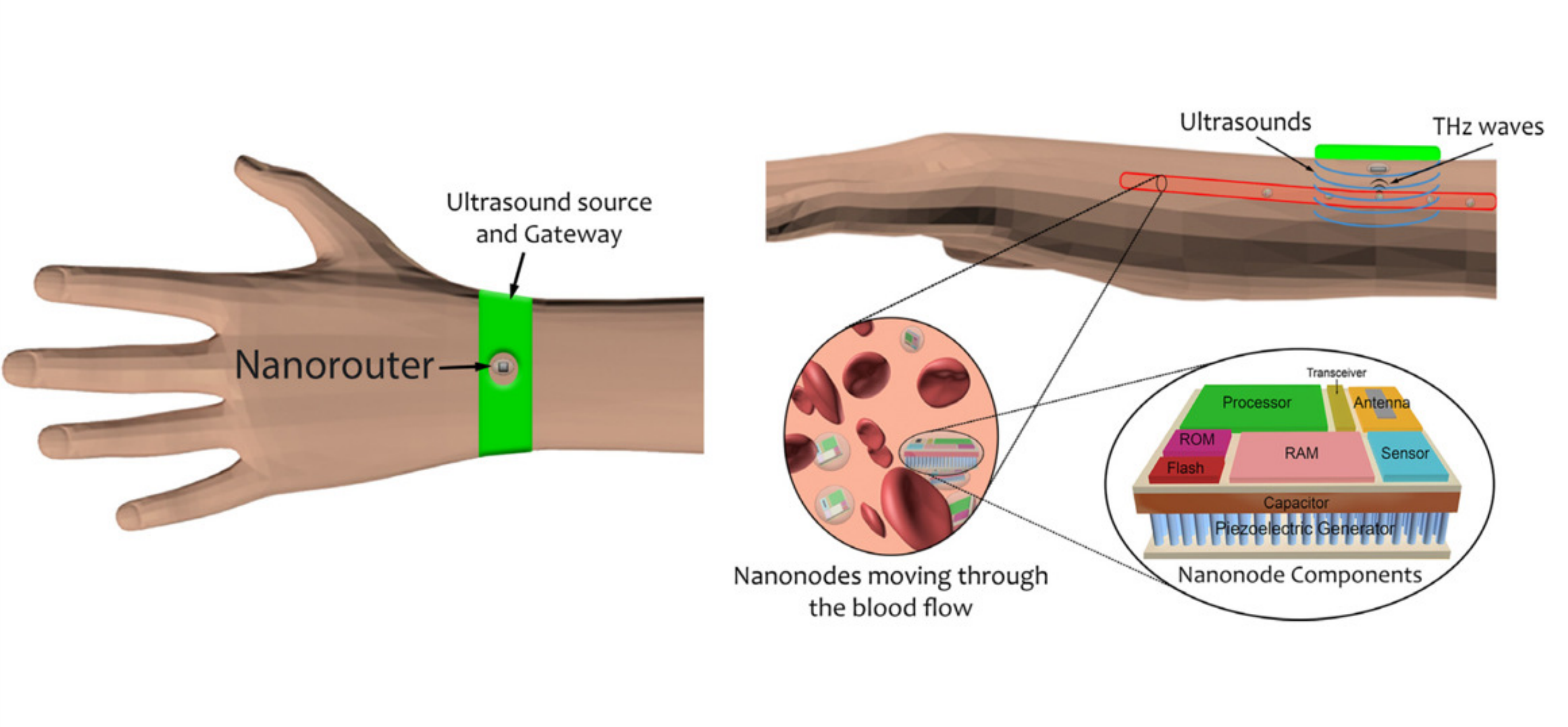}
		\caption{The proposed scheme in \cite{Canovas-Carrasco2018}.}
		\label{f_EM_2}
	\end{subfigure}
	\caption{Two nanonetworks schemes that adopt electromagnetic paradigm as their in-body and body-area communication method.}
	\label{f_EM}
\end{figure}

Some groups specified the communication paradigm with corresponding interface either using EM paradigm or MC paradigm. The authors of \cite{Islam} proposed a network deployment tailored for coronary heart disease monitoring, which is shown in Fig \ref{f_EM_1}. 
The network consists of two major components: Nanodevice-embedded Drug Eluting Stents (nanoDESs) and Nano-macro Interface (NM). The nanoDESs are deployed to occluded regions of coronary arteries and responsible for measuring the arterial constriction, communicating relevant information, and controlling the release of any required drugs. NanoDESs use THz band to communicate with an interface which is inserted in the intercostal space of the rib cage of a Coronary Heart Disease (CHD) patient and acts as a gateway between the nanonetworks and the macro-world. Another example that chooses THz communication is presented in \cite{Canovas-Carrasco2018}.
It proposed a nanoscale communication network consisting of nanonodes circulating in bloodstream and a nanorouter implanted between epidermis and dermis in hand skin, illustrated in Fig. \ref{f_EM_2}. The nanonodes in blood vessels collect healthy parameters and exchange data with the nanorouter using THz band only when they approach the nanorouter. In this way, the relatively short distance between nanonodes and the nanorouter minimizes the negative impact of path loss.
Subsequently, the nanorouter transmits the received information also in THz band to a gateway wristband that relays the healthy data to external devices or the Internet via traditional communication methods. 

As for MC paradigm, authors in \cite{Nakano2014} implemented artificially synthesized materials (ARTs) as an interface. In their wet laboratory experiments, the ART contains pHrodo molecules which are a kind of fluorescent dyes that are almost non-fluorescent under neutral solutions while fluorescent in acidic solutions. Therefore, conducting fluorescence microscopy observations and measuring fluorescent intensity can tell us the information inside our body. 

Apparently, all the above schemes can enable the connection between the In-Body Network and the Body-Area Network using electromagnetic paradigm or molecular paradigm, but there are some factors making them less practical. First, the nanonodes in \cite{Canovas-Carrasco2018} and nanoDESs in \cite{Islam} are non-biological and may intervene other physiological activities, as the nanonodes need to be injected into blood vessels or enter the human body through drinking a solution containing them, and the nanoDESs are even required to be surgically placed into body. 
Moreover, the injection or insertion of numerous nanonodes into the human body may not be accepted by the public, and some countries have published national laws to strictly regulate the production and marketing of such devices \cite{DRESSLER}.
Meanwhile, how to recycle these nanonodes is also a problem. Second, with regard to the method in \cite{Nakano2014}, the need of externally devices, fluorescent microscope, makes the method too complicated to implement for ordinary being. Furthermore, the fluorescent intensity information has to be transformed to electromagnetic form for the following transmission to the Internet. 

\begin{figure}[!htpb]
	\centering
	
	\begin{subfigure}{0.5\textwidth}
	\centering
	\includegraphics[width=0.4\textwidth]{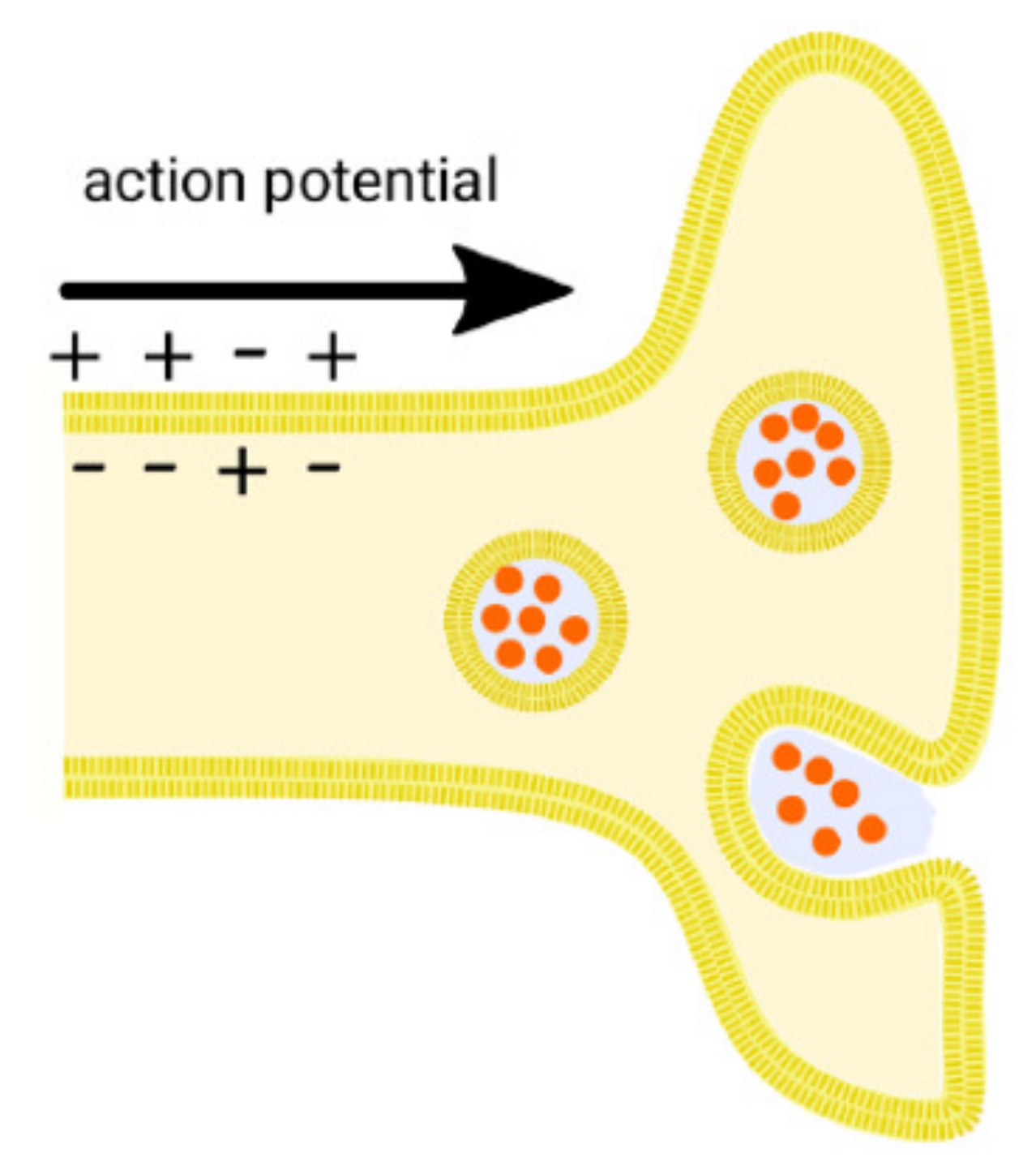}
	\caption{Transmission process of signaling particles}
	\label{f_NT_tx}
	\end{subfigure}
	
	\begin{subfigure}{0.5\textwidth}
	\centering
	\includegraphics[width=0.5\textwidth]{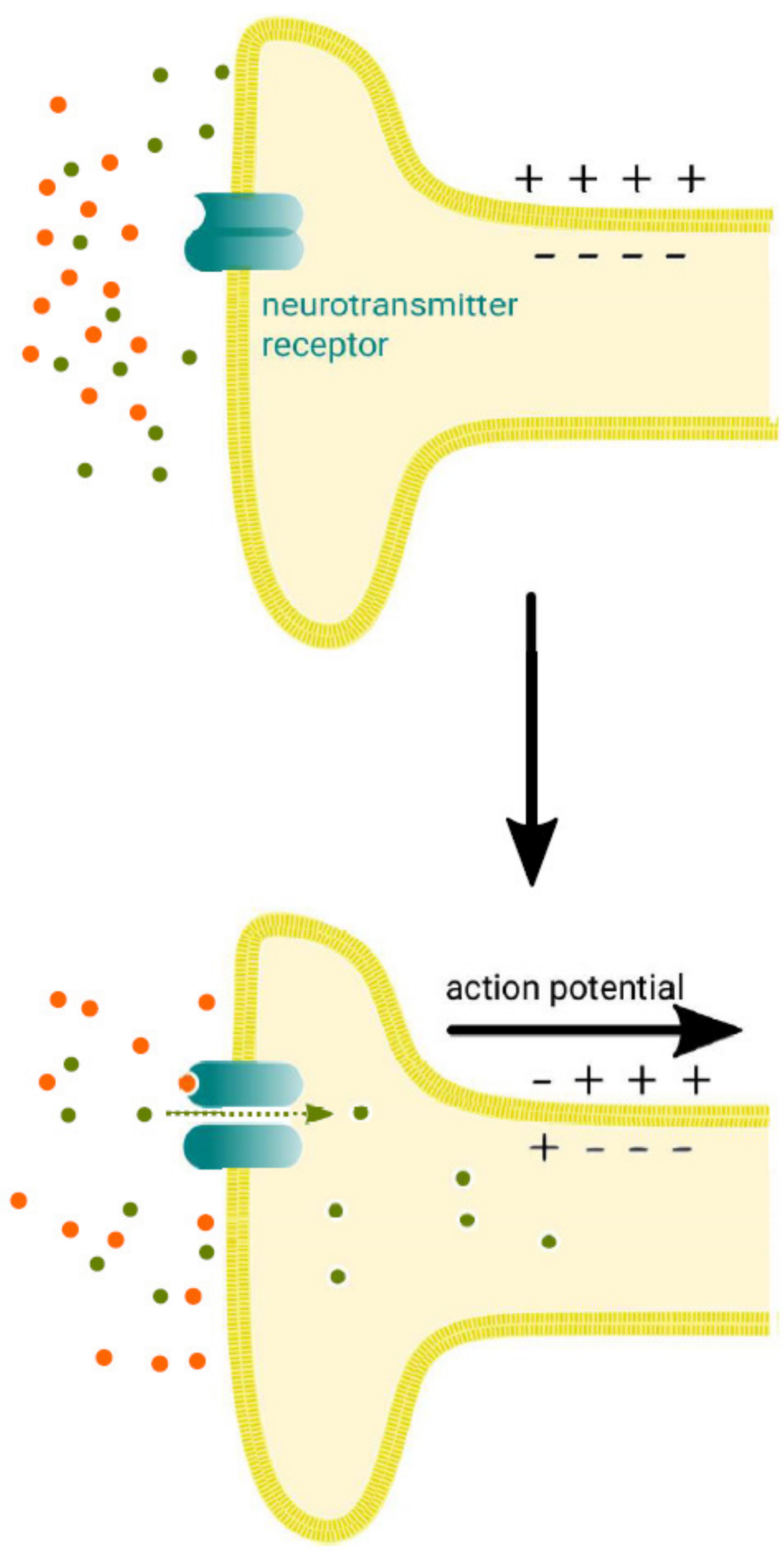}
	\caption{Detection process of signaling particles}
	\label{f_NT_rx}
	\end{subfigure}
	
	\caption{The transmission and detection of neurotransmitters. The red molecules are signaling neurotransmitters enclosed by roundshaped vesicles, and the green molecules are ion molecules who can result in a depolarization of the cell membrane \cite{Soeldner2019}.}
	\label{f_NT}
	\end{figure}

The nanoscale is the natural domain of molecules, proteins, DNA, organelles, and major components of cells \cite{EM2010}, \cite{Nelson2008}. \cite{Soeldner2019} investigated three kinds of possible signaling particles 
and discussed their corresponding biological building blocks to serve as transmitters and receivers for MC. A physiological process that happens naturally is the neurotransmitters transmission between presynaptic part and postsynaptic terminal, which is depicted in Fig. \ref{f_NT}.
In response to an excitation of a nerve fiber, the generated action potential moves along the presynaptic part and triggers the release of neurotransmitters (signaling particles) contained in vesicles. The released information molecules diffuse in the environment, and they can bind to the ion channel located at the membrane of postsynaptic terminal. Then, the binded ion channel becomes permeable to some ions, which the ion influx finally leads to a depolarization of the cell membrane that propagates subsequently as a new action potential along the cell \cite{Soeldner2019}, \cite{Clarke1999}. Undoubtedly, the neurotransmitter delivery establishes a MC link and is much more biological, biocompatible, and less invasive than nanonetworks systems consisting of nanonodes and using electromagnetic paradigm, since spontaneously existed molecular paradigms eliminate the risk of injection or intake of nano devices. In other words, the molecular paradigm makes up for the drawback of \cite{Islam,Canovas-Carrasco2018}. 

\begin{figure*}
	\centering
	\includegraphics[width=0.85\textwidth]{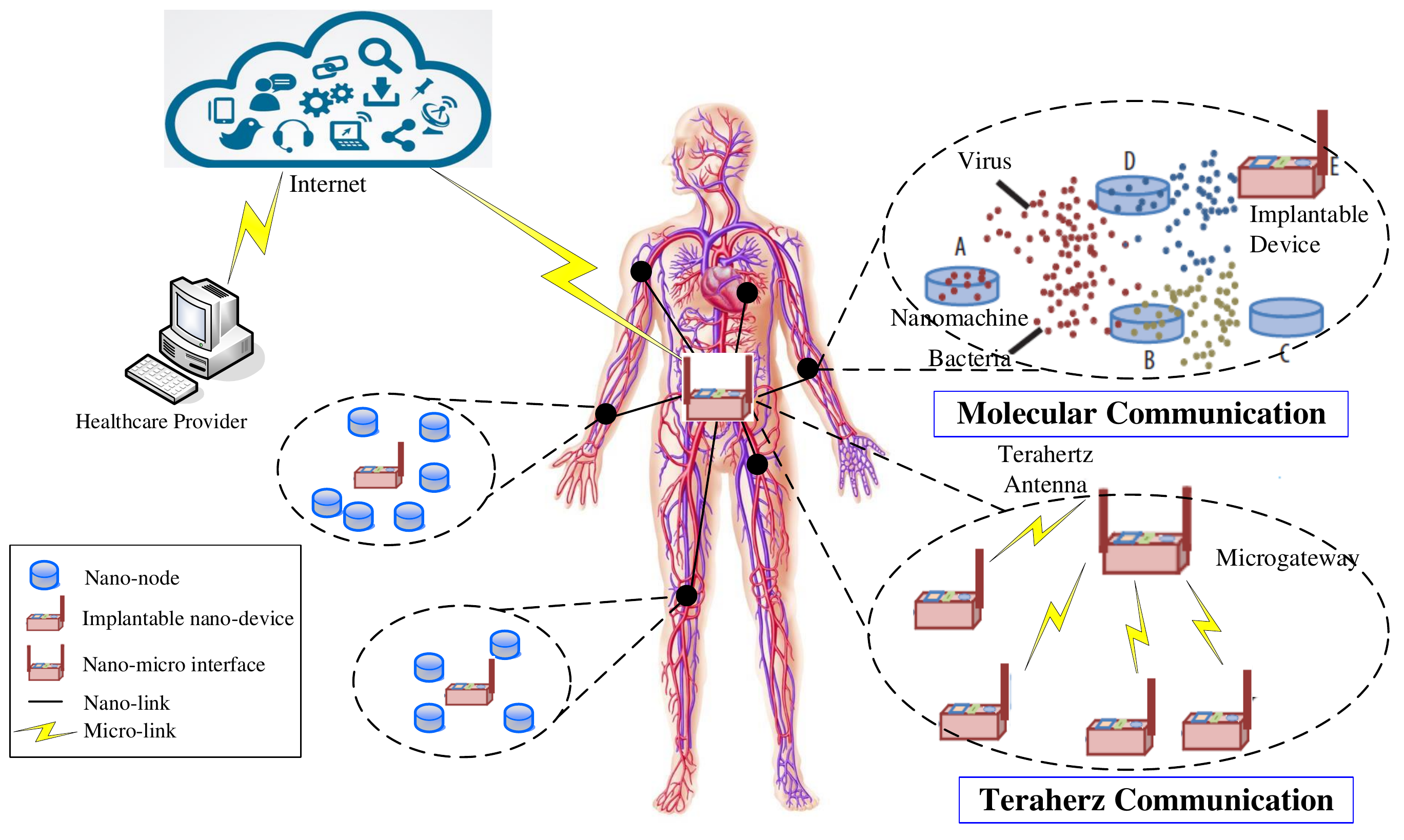}
	\caption{The sketch of the proposed nano communication network.}
	\label{f_hybrid_com}
\end{figure*}

Moreover, the implementation in \cite{abbasi2018controlled} further demonstrates the feasibility of that some physiological processes can be interpreted as MC systems. In MC, the information is generally modulated by molecules' concentration, while the information is usually transmitted outside the human body via electromagnetic waves, so a chemical concentration/electromagnetic wave convertor or interface is needed. Fortunately, some nanonodes with chemical nanosensors being embedded on the CNTs or GNRs are able to take this responsibility \cite{roman2004single,lazar2013adsorption,georgakilas2012functionalization}. The mechanism is that some specific type of molecules can be absorbed on the top of CNTs and GNRs, thus resulting in a locally change in the number of electrons moving through the carbon lattice and generating an electrical signal \cite{EM2010}. So far, the discussed advantages brougth by MC and electromagnetic communication provide the opportunity and open a door to propose a hybrid communication for nanonetworks systems.
 

\subsection{Hybrid nanonetworks Communication and Enabling Technologies}

Based on the opportunities offered by molecular paradigm and electromagnetic paradigm, we propose a hybrid communication that combines molecuar paradigm and electromagnetic paradigm for nanonetworks systems, which is shown in Fig. \ref{f_hybrid_com}. 

In the proposed hybrid communication network, the MC is utilized in the human body because it shows a superiority over other communication schemes
in terms of biocompatibility and noninvasiveness.
The blue nano-node in Fig. \ref{f_hybrid_com} refers to a MC system, and MC systems are grouped to constitute a molecular nanonetwork who is only responsible for a certain area. The molecular nanonetworks are either made up of multiple MC transmitters and receivers or a MC transmitter, MC receiver, and multiple transceivers that play the role of relaying. A biological transmitter
first collects health parameters, and then 
modulates and transmits the collected information among the molecular nanonetworks. In order to successfully delivery the information to the outside of the human body, a graphene based nano-device is implanted into the human body. This device is mainly made up of a chemical nanosensor, a transceiver, and the battery. The embedded chemical nanosensor is capable of detecting the concentration information coming from the molecular nanonetworks, and converts it to an electrical signal. The THz electromagnetic signal is further transmitted to a nano-micro interface. This interface can either be a dermal display device \cite{freitas1999nanomedicine} or a gateway to connect with the Internet. The nano-micro interface is usually equipped with two kinds of antennas: THz antenna and micro/macro antenna. The proposed hybrid communication architecture not only tries its best to avoid using non-biological nanonodes inside the body but also makes in-body healthy parameters easily be detected outside.

There are several enabling technologies to enhance the feasibility of the proposed hybrid communication. First, the molecular nanonetworks have been well studied (see Sec. \ref{MC_ber_bep}) \cite{arman2015mult,arman2015amplify,wang2015relay,mc_oc}. Different relaying or multi-hop schemes have been proposed and their performance are theoretically and numerically analysed, which demonstrate the effectiveness of communication distance extension and communication reliability improvement.
Then, the in-vivo THz communication including the channel modelling, modulation methods, and channel capacity has been studied (see Sec. \ref{THz_channel_model}) \cite{zhang2017analytical,zhang2018experimental,abba2016}. The conducted research not only helps us understand the impact of human tissue on signal propagation but also assists researchers to estimate the received signal level which is a key indicator for the further information transmission.

\subsection{Challenges and Open Issues}
The integration of molecular paradigm and electromagnetic paradigm will boost the application of medical monitoring. However, this combination also imposes some challenges. The first concern comes from MC. As the existing biological system descried in Sec. \ref{requirements_opportunities}, the signal delivery process involves the subjects of electrochemistry, neuroscience, and biology.
Hence, the highly interdisciplinary technical knowledge and tools required to analyse puts forward higher requirements for researchers. At the same time, although \cite{Soeldner2019} presents some possible MC systems occurring in the human body, biological transmitters and receivers are still required to be modified to be tailored to various application needs. Given the synthetic MC is in its infancy, the design, analysis, and implementation of synthetic MC also inherently require a multidisciplinary approach \cite{Soeldner2019}. Another concern is the choice of signaling particles. We hope the signaling particle is a kind of neutral or intermediate substance. It should be easily detectable, non-toxic to the human body, and cannot intervene other biological processes. 
Meanwhile, the selected signaling molecular is supposed to be reversible, which means it can be recycled and used for repeated transmission. The finding of an ideal candidate needs a further study of various human physiological processes.

From the THz communication perspective, the obtained channel parameters may not fit everyone because channel characteristics for intra-body nanonetworks may vary with health conditions and from person to person \cite{rizwan2018review}. Thereby, further investigation of channel modelling is needed.

\section{Security in nanonetworks: Progress \& Open Issues}

The fundamental goals of security schemes are to ensure confidentiality, integrity, and availability of the data exchanged between the legitimate nano nodes. This section summarizes the works which have attempted to address/raise the security challenges faced by (EM/MC based) nano networks, and provides authors' vision about the nature of the open security issues and their potential solutions. But, before we outline our vision of the security in nano networks, it is imperative to quickly review and summarize the main ingredients of security in traditional wireless networks.

\subsection{Security in traditional wireless networks}
In traditional wireless networks, communication between legitimate nodes is prone to active and passive attacks by adversaries, due to the broadcast nature of the wireless medium. The literature have considered various kinds of attacks, e.g., impersonation attack, Sybil attack, Replay attack, Sinkhole attack, jamming, man-in-the-middle attack, denial of service attack, eavesdropping attacks, selfish/malicious relays in cooperative communication systems etc., and their potential (cryptography based) solutions. More recently, researchers have started to develop various security solutions at physical layer by exploiting the unique characteristics of the physical/wireless medium. Some of the most significant problems in physical layer security include intrusion detection/authentication, shared secret key generation, secrecy capacity maximization (for a wiretap channel), artificial noise generation, design of friendly jammers (in a cooperative communication system) etc. 

Keeping this context in mind, we evaluate the answer to the following question: do the aforementioned security solutions hold for the nano-scale communication? The answer is in negation for MC based nano networks because information exchange by using molecules instead of EM waves as carriers is a different regime altogether. On the other hand, we find that for EM based nano networks, operating at THz frequencies, some of the aforementioned concepts (if not the solutions) are still meaningful. 

\subsection{Security in EM based Nano networks}
As explained earlier, EM based nano-scale communication at THz frequencies is a relatively new phenomena which has garnered much interest recently only because the device fabrication techniques are now approaching to the level of miniaturization needed to fabricate nano transmitters and receivers (e.g., graphene based nano antennae etc.). This implies that the EM based nano-scale communication, being an extremely short-ranged communication regime, is still at risk of passive and active attacks by adversaries in the close vicinity. Nevertheless, due to THz band communication being in its infancy, not much works are available in the open literature which investigate the security issues faced by THz systems. On the contrary, THz waves have a long-standing history of being used for imaging, sensing etc. for security purposes \cite{Federici:SST:2005}. However, THz based imaging systems are not the focus of this survey article.

The survey articles \cite{Dressler:ICC:2012}, \cite{Dressler:NanoCommNet:2012} review some of the fundamental security mechanisms for THz systems and conclude that the traditional crypto based mechanisms could be ported to THz systems, but they need to be light weighted due to limited processing capabilities of the THz devices. The so-called BANA protocol proposed by Shi et. al. in \cite{Shi:JSAC:2013} addresses the security needs of the micro-macro link of a body area network. In \cite{Mahboob:Access:2017}, the authors consider a scenario where an on-body nano device communicates with inside-body nano device, while a malicious node attempts to send malicious/harmful data to the inside-body node. To this end, the authors utilize the measured pathloss as the fingerprint of transmit nano device to do the authentication at the physical layer. \cite{Punthawanunt:JBiosenBioElect:2016} presents the device layout which consists of a micro-ring transceiver and a graphene based panda ring-resonator. The molecules are trapped in a whispering gallery mode, the polarized light is transceived and this device which could be used as a molecular RFID system.

\subsection{Security in Molecular based Nanonetworks}
For MC networks, the traditional crypto based methods need to be replaced by the so-called biochemical crypto techniques whereby attacks as well as countermeasures are all defined by the chemical reactions between the molecules \cite{Dressler:NanoCommNet:2012}, \cite{Dressler:ICC:2012}. Various bio-inspired approaches are proposed in \cite{Loscri:TNBS:2014} to secure MC systems and different attacks are classified according to the (five) different layers of MC system in Table \ref{fig:mc}. From the table, we can see that besides the classical attacks numerous other novel attacks are possible.

\begin{table}[]
    \centering
    \caption{Potential security attacks at each layer \cite{Loscri:TNBS:2014}.}
    \begin{tabular}{p{100pt}|p{110pt}}
         \hline
         Layer & Type of Attack \\ \hline
         Molecular Transport Layer & Unfairness, Desynchronization \\ \hline
         Molecular Network Layer & Exhaustion of packets storage. Flooding attacks \\ \hline
         Molecular Link Layer & Collision, Unfairness \\ \hline
         Signaling sublayer & Jamming, Misuse of Replication functionality \\ \hline
         Bio-nanomachine sublayer & Jamming, Tampering \\
         \hline
    \end{tabular}
    \label{fig:mc}
\end{table}

Two kinds of attacks are discussed in \cite{Giaretta:TIFS:2016}, which are \textbf{blackhole attack} where malicious bionano things attract the other bionano things towards itself (by emitting chemo-attractants) preventing them from their task of localization, and \textbf{sentry attacks} where malicious bionano things in the vicinity of the target cells emit chemo-repellents not letting the legitimate bionano things reach their target. Reference \cite{Guo:CommL:2016} and \cite{Islam:arxiv:2017} consider the situations that eavesdropper appears and causes troubles; furthermore, the solutions are also discussed and evaluated. Additionally, in vesicle based molecular transport, vesicles act like keys in MC networks and thus inherently help the cause of secure communication. Recently, researchers from cryptography have extensively work on DNA inspired cryptography \cite{Clelland:Nature:1999} \cite{Leier:BioSys:2000} \cite{Gehani:LNCS:2003} \cite{Monica:ICCOMM:2010},  the crux of which is that DNA computing is a computationally hard problem of biological origin, just as Heisenberg's uncertainty principle is a hard problem of physics origin; thus, this could be applied to cryptography purposes.

\subsection{Security in Hybrid nano-scale communication}
The proposed hybrid nano-scale communication systems could either switch between EM/MC mode from one leg to another, or, from one time-slot to another. For the former scenario, the aforementioned details of the security challenges and potential solutions hold as is (as the nano network under consideration will be either MC based or EM based at a given leg and will be secured accordingly). For the latter scenario, authors envision that one could develop a systematic approach that optimally switches between MC mode or EM mode depending upon the security requirement of a given application and/or extent of hostility (or trustworthiness) of the environment nano network is operating inside.

\section{Conclusion and Recommendations}
With the development of the novel manufacturing techniques, the size of the sensors or machines can be made as small as micro scale, even smaller to nano-level; however, the realization of the nanonetworks is still very challenging. It is generally believed that it is hard to fulfill the ultimate goal only by one individual communication method; thus, the hybrid communication method should be fully studied, and it is believed that the hybrid communication is firstly studied in this survey. By combining the EM and molecular communication, the drawback of short communication distance for EM and huge latency for molecular communication can be solved simultaneously. However, how to combine both paradigms seamlessly still need effort. In this survey, the connectivity of EM and molecular methods is investigated, and it shows promising potential; however, it is far from achievement. First, the interfaces for both methods are difficult to design because of the different information throughput; thus, how to design an interface which can balance both network throughput is of great interest. Second, there is no uniform simulation platform for the hybrid communication because of the different network realization pattern. Even for different molecular methods, there are different analytic models. Because of the diversity, it is hard to unify all the communication methods. Third, how to deal with the conflict of the nano-devices is also important because of the different latency of both paradigms. It is believed that the performance of the channel would change with the composition of the medium; however, the communication process of the molecular method would change the channel composition. Therefore, it would be of great importance to study the effect of the molecular communication on the EM channel.

Besides the challenges and the corresponding future work presented above, there are still other research directions which can be summarised as follows:

\subsection{Investigations of the Novel Materials}
From the previous experiences, the discovery of a novel material would make the development of the engineering leaping forward by a huge gap. Taking the investigation of graphene for an example, the study of its characteristics solidates the concept of nano EM communication with the application of THz technology, which sets up the foundation of the current work. Recently, the study on the Pervoskite enables the design of the Terahertz Antenna which brings a bright future of the short-range body-centric communication {\cite{green2014the}}. It is generally believed by the author that the researches on the bio-materials would rocket the development of the truly nano-devices, enabling the realization of nanonetworks. Also the new material would enable the transfer from different signal types which would make the conversion between different communication methods possible. Additionally, we all believe that the more discoveries on the bio-materials, the closer the nanonetworks comes into reality. 

\subsection{Integration Techniques of Diverse Communication Methods}
Currently, every communication methods seems well-developed in their own fields; however, the transformation between each others is still at its initial stage, especially for the nano-devices. In the survey, such problems are discussed in Section VII but only in the shallow phase. The researches on the interfaces between the different communication methods should be deeply investigated which are apparently not sufficient right now, not only from the material perspective but also on the devices level. The same problems occur for the interfaces of macro-and-micro scenarios. It is believed that if such problems are tackled, the major problems in nanonetworks would be solved.

\subsection{Development of the Generalized Platform}
The ultimate goal would be the integration of all communication methods in nanonetworks. However, the studies on the communication methods are independent with each other. Moreover, the interactions between two methods are missing. For example, the emission of the molecules would change the channel environment for EM ones. And at the same time, it is clear from the previous studies that the refined models are necessary like the nerve system and skin. Therefore, the platform to simulate the interaction between the environment and the devices should be further studied to make sure the whole network act normal. Although in IEEE P1906.1, different schemes are studied according to EM standard based on NS-3 Platform; but the general platforms to simulate hybrid communication with every part under considerations are still absent and not well studied.

\subsection{Introduction of Big Data Analysis Techniques}
The integration of big data techniques with the nanonetworks would be a hot topic because of the nature of nanonetworks that numerous stakeholders are involved in data generation and management. From the perspective of the data analysts, the study is still missing. First, the standardization of the data-format and protocols should be set up while a unified data schema should be put forward and adopted by the whole network investigators. Most importantly, new powerful analytic tools should be developed because the amount of the data we will face would go up hugely.


%

%
%
%

\ifCLASSOPTIONcaptionsoff
  \newpage
\fi

\end{document}